\documentclass[]{aa}

\usepackage{graphicx}
\usepackage{txfonts}
\begin{document}

\titlerunning{Long term NGC 3516 broad line profile variability}
\authorrunning{L. \v C. Popovi\'c et al.}

   \title{Long-term optical spectral monitoring of a changing-look active galactic nucleus NGC 3516 - II. Broad-line profile variability }

   \author{Luka \v C. Popovi\'c
   \inst{1,2} 
   \and
   Dragana Ili\'c
   \inst{2,3}
   \and 
   Alexander  Burenkov
    \inst{4}
   \and
   Victor Manuel Pati\~no Álvarez
    \inst{5,6} 
   \and 
   Sladjana Mar\v ceta-Mandi\'c
   \inst{1,2} 
     \and
   Jelena Kova\v cevi\'c - Doj\v cinovi\'c
   \inst{1}
   \and
   Elena Shablovinskaya
   \inst{4} 
   \and
   Andjelka B. Kova\v cevi\'c
    \inst{2,7}
    \and
   Paola Marziani
   \inst{8} 
   \and 
   Vahram Chavushyan
   \inst{5,9} 
   \and
   Jian-Min Wang
   \inst{7,10,11}
   \and
   Yan-Rong Li
   \inst{7} 
   \and
   Evencio G. Mediavilla\inst{12,13} 
   }

    \institute{Astronomical Observatory, Volgina 7, 11000 Belgrade, Serbia \\
    \email{lpopovic@aob.rs}
    \and
    Department of Astronomy, University of Belgrade - Faculty of Mathematics, Studentski trg 16, 11000 Belgrade
    \and
    Humboldt Research Fellow, Hamburger Sternwarte, Universit{\"a}t Hamburg, Gojenbergsweg 112, 21029 Hamburg, Germany
      \and
    Special Astrophysical Observatory of the Russian AS, Nizhnij Arkhyz, Karachaevo-Cherkesia 369167, Russia 
       \and 
  Instituto Nacional de Astrofi\'isica, \'Optica y Electro\'onica, Luis Enrique Erro 1, Tonantzintla, Puebla 72840, Me\'exico 
  \and
  Max-Planck-Institut fuer Radioastronomie, Auf dem Huegel 69, D-53121 Bonn, Germany
        \and  
 Key Laboratory for Particle Astrophysics, Institute of High Energy Physics,
Chinese Academy of Sciences, 19B Yuquan Road, Beijing 100049, China
 \and 
    INAF Osservatorio Astronomico di Padova, Vicolo dell'Osservatorio 5, Padova, Italy
  \and 
  Center for Astrophysics | Harvard \& Smithsonian, 60 Garden Street, Cambridge, MA 02138, USA
  \and
School of Astronomy and Space Sciences, University of Chinese Academy of Sciences, 
19A Yuquan road, Beijing 100049, China
  \and 
National Astronomical Observatory of China, 20A Datun Road, Beijing 100020, China
 Beijing China
   \and 
    Instituto de Astrof\'isica de Canarias, V\'a L\'actea S/N, La Laguna 38200, Tenerife, Spain
    \and 
    Departamento de Astrof\'isica, Universidad de la Laguna, La Laguna 38200, Tenerife, Spain}
   
   \date{Received March 25, 2022; accepted June 16, 2022}

% \abstract{}{}{}{}{}
% 5 {} token are mandatory

  \abstract
   {We analyze the broad H$\beta$ line profile variability of the "changing look" active galactic nucleus (CL-AGN) NGC 3516 over a long period of  25 years (from 1996 to 2021). The observed change in the broad line profile may indicate a change in the geometry of the broad line region (BLR). The main objective is to follow and understand the change in the BLR over a long period and its connection with the CL mechanism.}
   {Using spectral line profiles, we aim to explore changes in the kinematics and dimensions of the BLR in NGC 3516.  We consider two possible scenarios, i.e. changes in the broad-line emission are caused by a decrease of ionization continuum emission or by the BLR obscuration by outer dusty regions. With this investigation we aim to clarify the CL mechanism  of this AGN. }
   { We analyze the spectral band around the  H$\beta$ line as well as the broad  H$\beta$ line parameters, and how they   change in time. We model the broad-line profiles assuming that there is an emission from the accretion disc superposed with an emission from a surrounding region that is outside the disc.}
   {We find that in the Type 1 activity phase  (i.e. when the strong broad emission lines are observed), the BLR is very complex. There is a clear disc-like BLR that contributes to the broad line wings and an additional intermediate line region (ILR) that contributes to the line core. In the high activity phase, the ILR emission is close to the center of the line (in some cases slightly shifted to the red), whereas in the low activity phase  (i.e., Type 2 phase), the ILR component has a significant shift to the blue, indicating an outflow.  }
   { We propose that the changing look mechanism in NGC 3516  is rather connected with the intrinsic effects than with an outer obscuring region. It may still be possible that the dust has an important role in the low activity phase when it is coming inside of the BLR, making a dusty BLR. In this way, it causes a decrease in the ionization and recombination rates.}

% when it comes inside the BLR, i plunge into ili emerse into
   
   \keywords{(galaxies:) quasars: individual: NGC 3516 -- lines: profiles}

   \maketitle
%
%-------------------------------------------------------------------

\section{Introduction}

The broad emission lines (with Full Width at Half Maximum - FWHM around several 1000 km s$^{-1}$), which are emitted from the broad line region (BLR) can be observed in, so called  Type 1 active galactic nuclei (AGNs). According to the Unified model, the broad lines (emitted from the BLR) can be observed when an AGN has an adequate orientation \citep[see][]{an93,up95} with respect to the observer. In this scenario, the BLR is present in all AGNs, but the emission from BLR can only be detected when it is not covered by the dusty torus.
In principle, the difference between Type 1 and 2 AGNs is caused by obscuring of the dust that is assumed to be in a shape of a torus that is co-planar with the accretion disc \citep[][]{up95}. However, the dust distribution can be more complex, and dust can be also present in the form of  polar cones in AGNs \citep[see][]{st19}. 

In the case of Type 1 AGN, one can see emission of the gas that is closer to the  central supermassive black hole (SMBH) and 
the broad line widths are broadened due to the rotation caused by gravitation of the SMBH, therefore they are often used for the SMBH mass measurements \citep[see review][and references therein]{pop20}.
However, other effects, such as inflows and outflows could affect the width of broad lines \citep[see e.q.][]{le13,pop19,hu20}. The broad line shapes can be used to explore the BLR kinematics and emission gas motion that is close to the SMBH.
The broad line variability is caused by the variability of the ionized gas source, but dust can play an important role because it can obscure a portion or the entire BLR \citep[see, for example,][]{gh18}. The variation of the broad lines intensity is mostly caused by the variability of the ionization continuum, but the change in line shapes, widths, and shifts is most likely caused by the change in the BLR geometry \citep[see e.g.][etc.]{sh10,po11,du18,br20,ba22,ch23}. Both of these changes can give more information about the nature of the AGNs.

Additionally, in a number of AGNs, the broad component in emission lines has been disappearing for a certain period of time, during which they only show a narrow component like in Type 2 AGNs  \citep[see e.g.,][for early works]{ly84,ko85}.
These AGNs are classified as changing look AGNs (CL-AGNs). This phenomenon of CL-AGNs may be caused by several reasons, among which are very often mentioned: a) existence of an obscuring material, which is able to cover the BLR and obscure the broad line component
\citep[e.g.][]{el12}; b) changing in the rate of accretion, i.e. from time to time, there is lack of accreting material \citep[e.g.][]{st18}, that may be caused by different reasons 
\citep[e.g. tidal disruption events][]{er95}. Also, the combination of these two effect as well as some additional effects can be responsible for CL-AGN phenomena \citep[see e.g.][etc.]{no16,no18,gu20,wb20}.

The question of the CL AGN nature is one of the most interesting in the AGN investigation \citep[see e.g.][]{bi05,ma16,no18,ma18,ma19,sn20,sn22}, since it indicates that the difference between Type 1 and Type 2 AGNs may not be caused only by their orientation. Several tens of AGNs have been observed as changing  look AGN up to now 
\citep[see e.g.][]{ya18,ho20,ho22,ko22},  and more are to come with ongoing and coming time-domain surveys \citep[see e.g.][]{ss2021}. 
% https://ui.adsabs.harvard.edu/abs/2021AJ....162..206S/abstract
One of the characteristics of the CL-AGN is that the broad line appearance is connected with the blue continuum in the most of the observed CL-AGNs \citep[see][]{ho22}.
%(see e.g. Hon et al. 2022)%https://arxiv.org/pdf/2202.04851.pdf

The variability of the broad line flux and shape may help to clarify the nature of the CL-AGNs, especially in the case of long-term monitoring of this type of AGNs. Specifically, the line shape variability is connected with the BLR geometry and can indicate changes in the BLR geometry, i.e., if it remains the same or, contrary to that, if it is significantly changed during the transition between two phases of activity.
%Nadam se da je sada jasnije
In Shapovalova et al. (2019, hereafter Paper I) we published the variability in the flux of the continuum and emission lines of the active galaxy NGC 3516 during a long time period (1996--2018), in which we confirmed that this AGN is a CL-AGN. We found that the broad lines disappeared in period after 2014, with a 
 very weak broad line component observed in 2017, that supports the changing-look classification of this AGN. The phase of weak broad lines is extending after 2017, during which a number of coronal lines have been detected \citep[see][]{il20}. 
 
Galaxy NGC 3516 was one of the first observed active galaxies \citep[][]{se43}, and the strong variability of its broad emission lines  was firstly noted by  \citet{an68}.
%(Andrillat \& Souffrin, 1968).  
The galaxy is in the local universe (z$\sim$0.009) and is a  bright object (V$\sim$12.5 magnitude) which is the reason why this object was very often observed in spectral mode 
\citep[see][etc.]{so68,an71,co73,ad75,bo77,os77,cr86,co88,bo90,wa93,ly93,wa94,de10,de18,sh19,il20}.
In this paper we will explore the H$\beta$ broad line shape of NGC 3516 over a long period, in order to explore possible  changes in the BLR geometry during a transition phase  from Type 1 to Type 2 AGN.
To address the question of the CL nature of NGC 3516, we explore the changes in the broad line shapes in a long term period (from 1996 to 2019,  reported in Paper I), with additional observations from 2019 to 2021.  The aim of this paper is to investigate the H$\beta$ broad line profile observed in a long term period in order to find the nature of the CL mechanism in NGC 3516.

The paper is organized as follows: in section 2 we shortly  describe the observations and  the data reduction; in 
section 3 we model line profiles and outline the most important results of the broad line profile modeling. In \S 4 we discuss the obtained results in the frame of the CL nature of NGC 3516;
and finally in \S 5 we outline our conclusions.

\begin{figure*}[]
\centering
\includegraphics[width=6.5cm]{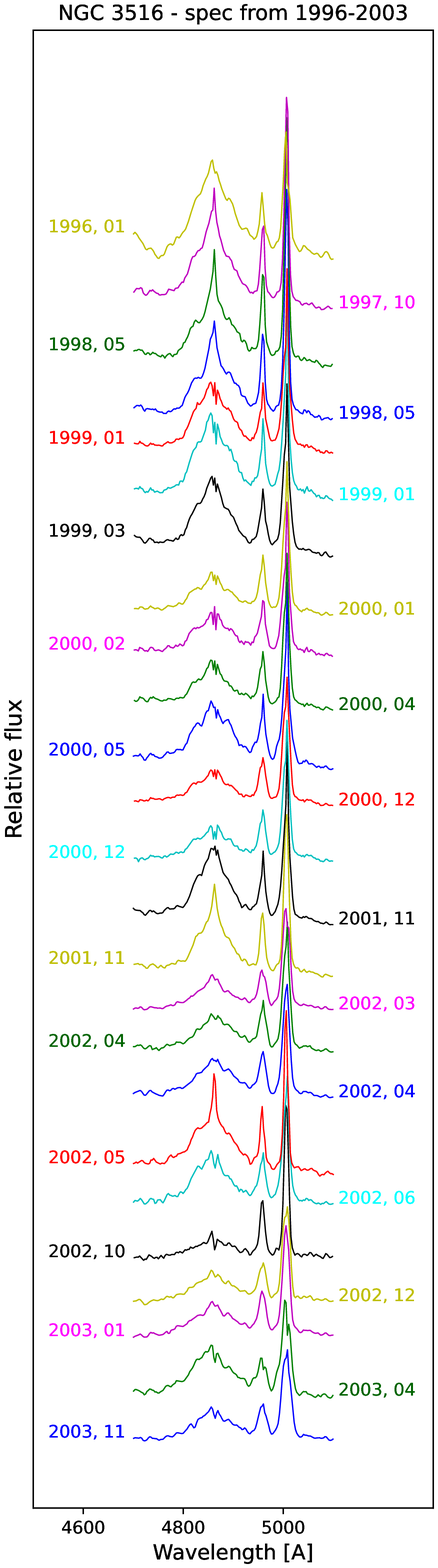}
\includegraphics[width=6.5cm]{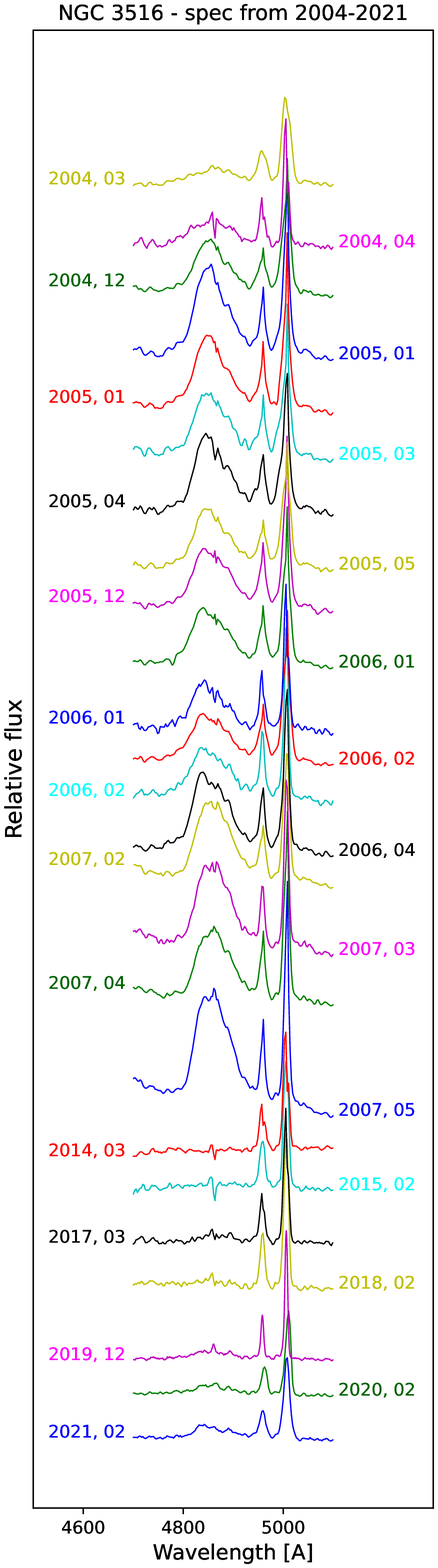}
\caption{The  H$\beta$ wavelength band of NGC 3516 observed from 1996 to 2021. The spectra are normalized to the [O III] line flux.}
\label{fig00}
\end{figure*}

\section{Observations and spectral analysis}

\subsection{Spectral observations}

Details about the observations, calibration of spectra, 
unification  of the spectral data and measurements of the spectral fluxes are given in our previous works \citep[see][and references therein]{sh04,sh08,sh10,sh12,sh13,sh16,sh17,sh19} 
and here we will not be repeated, but we will give some basic information about the observations and data reduction.

\begin{figure*}[]
\centering
\includegraphics[width=12cm]{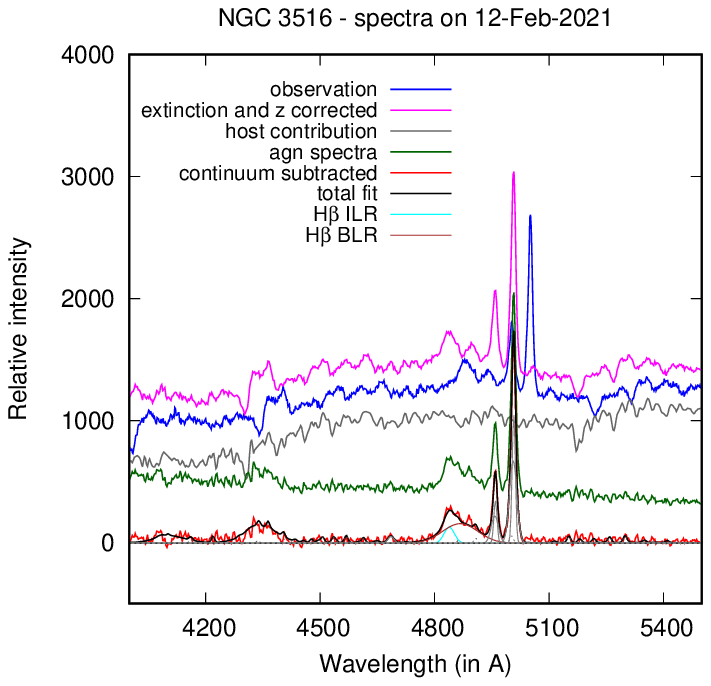}
\caption{Example of the procedure of the extracting broad H$\beta$ component for a spectrum with weak Balmer lines (from 12-Feb-2021). First,  the host galaxy contribution was subtracted from the observed spectrum. After that, the continuum has been estimated and subtracted and the broad and narrow lines were fitted (bottom spectrum).}
\label{fig-p}
\end{figure*}

NGC 3516 has been  monitored in a large period from  1996 to 2021. The observations from 1996 to 2019  were taken with the 6 m and 1 m telescopes of the SAO RAS, Russia, while with the INAOE's 2.1 m telescope of the "Guillermo Haro Observatory" (GHO) at Cananea, Sonora, Mexico the observations were taken in period 1998--2007 and 2020--2021. The description of observations and data reduction for these spectra were  described in detail in Paper I.  Note that  Paper I reported on data acquired until 2019, however, all spectra collected afterwards were processed in the same manner.

\subsection{Spectral  analysis}

To explore broad line profile from different epochs, we mostly used H$\beta$ line profile because of two reasons: first, the most spectra presented in Paper I are covering only the H$\beta$ wavelength band, and, second, the spectral line intensities are re-scaled to the [O III] lines which is close to the H$\beta$, therefore we avoid some possibility of deformation of the line intensity and shape farther in the spectra, where H$\alpha$ is located. However, we used H$\alpha$ broad line only in 2017 and 2019, since we show in \cite{sh19} that the H$\beta$ and H$\alpha$ profiles are the same in 2017.

To analyze the broad emission lines, first we calibrated the spectra according to the [O III] lines \citep[see in more details][]{sh19}. { However, we note that the [O III] lines could vary over a long period \citep[as e.g., in the case of NGC 5548, see][]{pet13}, especially in this case after the ionizing continuum luminosity decreased dramatically (in the period 2014-2018). In this study, we will mostly focus on broad line profiles where the effect of normalization on [O III] lines should not strongly affect the results. In the part where we compare changes in the broad line profile with broad line intensity, one should keep this in mind.}

\begin{figure*}[]
\centering
\includegraphics[angle=-90, width=0.89 \textwidth]{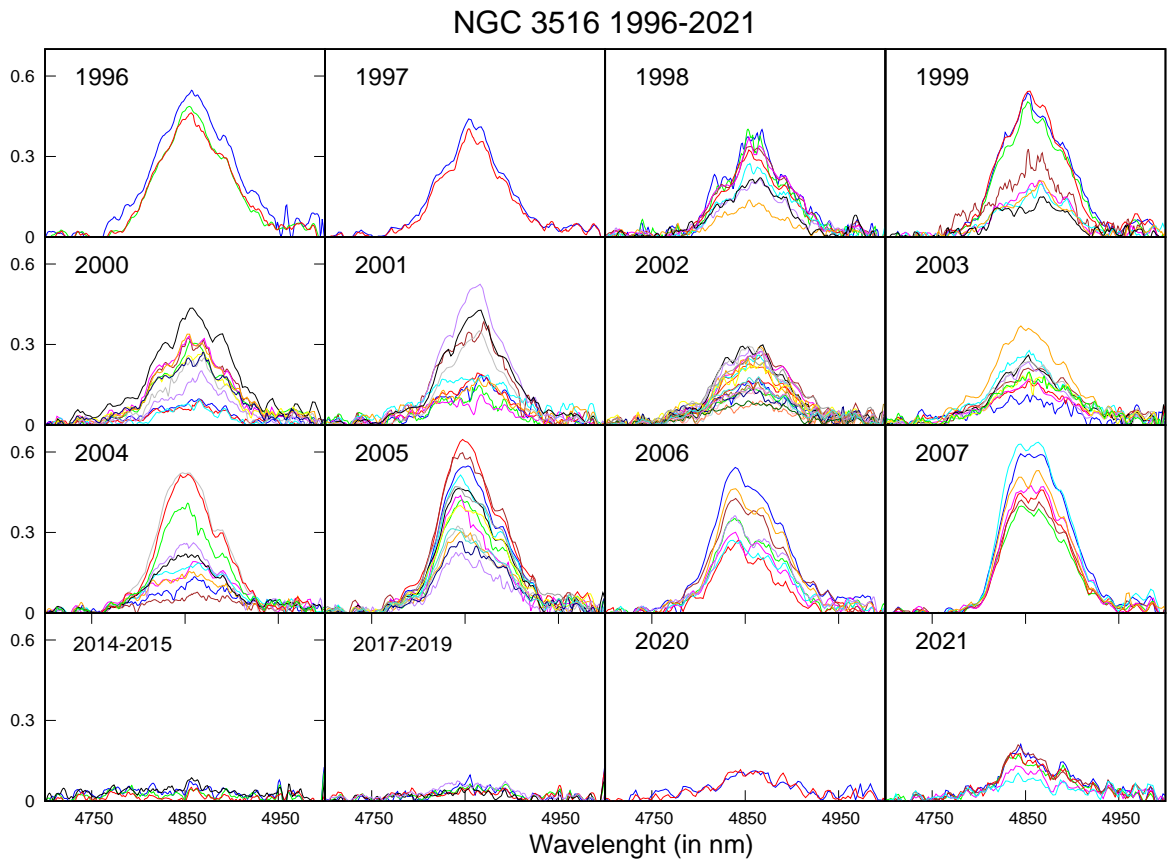}
\caption{All the  H$\beta$  broad profiles  observed during the monitoring period.  The extracted broad line profiles are grouped within calendar years. The broad H$\beta$ intensity is normalized to the [O III] lines (see Fig. \ref{fig00}). }
\label{fig01}
\end{figure*}

The H$\beta$ wavelength region observed in different epochs is presented in Fig. \ref{fig00}. To extract only the broad  H$\beta$ components, we performed the following steps (see Fig. \ref{fig-p}):
\begin{itemize}
\item  The reproduction and subtraction of  the host galaxy spectrum. This is very important for spectra obtained during the low activity phase, when broad component was very weak. 
\item  Fitting the H$\beta$ emission region, taking assumptions that broad line is composed from two Gaussian function, and narrow lines (narrow H$\beta$ and [O III] lines) have the same kinematics and other constrains \citep[see e.g.,][]{po04,di07,ko10}.
\item Subtraction of the continuum and narrow lines in order to have only broad line components from different epochs, as it shown in Fig. \ref{fig01}.
\item The broad H$\beta$ lines  
%from different epochs 
are normalized to unity, and the wavelength scale is converted into velocity scale ($v/c$) in order to compare the line parameters (FWHM and shift) and shapes from different epochs (see Fig. \ref{fig01a}) 
\end{itemize} 

Since H$\beta$ line is very weak in some epochs (e.g., in the period 2014–2019), we adopted for further analysis only the lines where the maximal intensity in broad H$\beta$ over noise in H$\beta$ is larger or approximately equal to 3. In this way, we rejected one observation from 2014. To estimate the error in fitting parameters (line widths, shifts, and asymmetries), we applied the Monte Carlo method for each spectrum. First, we measured the $\sigma$ noise in each spectrum, and then we used the line parameters obtained from the best fit to make the 100 mock spectra by adding random noise with the same level as measured in the observed spectrum. We took the 2$\sigma$ dispersion of the fitting parameters of 100 mock spectra as the fit uncertainty.

Additionally, for each year we calculated the mean and rms profile in order to see changes in the line profile (see Fig. \ref{fig01a}).

\section{Results}

The broad H$\beta$ line shapes from different epochs have been inspected (see Figs. \ref{fig00} and \ref{fig01}) and broad line parameters have been measured. We also fitted the broad line profiles with a two-component model  (accretion disc and surrounding region) and here we present the obtained results.

\subsection{Broad H$\beta$ line profiles}

The  change in the broad line shape can indicate  changes in the BLR geometry. In order to study this, we calculate the mean and rms profiles for each calendar year 
%from different epochs, 
during the monitoring period.
As it can be seen in Figs. \ref{fig00} and \ref{fig01} there are different states of AGN activity seen in NGC 3516 \citep[see also][]{sh19}.  Below we comment the behaviour of broad H$\beta$ line during the monitoring period from 1996 to 2021.

From the beginning of our observational campaign reported in \citet{sh19}, the H$\beta$ line was very intensive
(epoch 1996--1999). After that, a period of variable  broad line intensity is observed in 1999--2001, while in the period of 2002--2003, H$\beta$ was weak. After 2004 until 2007 the H$\beta$ becomes more intensive. The broad lines almost disappear in observations from 2014, and there is a very weak broad component  appearing after 2017 \citep[see][]{sh19,il20}.

\begin{figure*}[]
\centering
\includegraphics[angle=-90,width=0.89\textwidth]{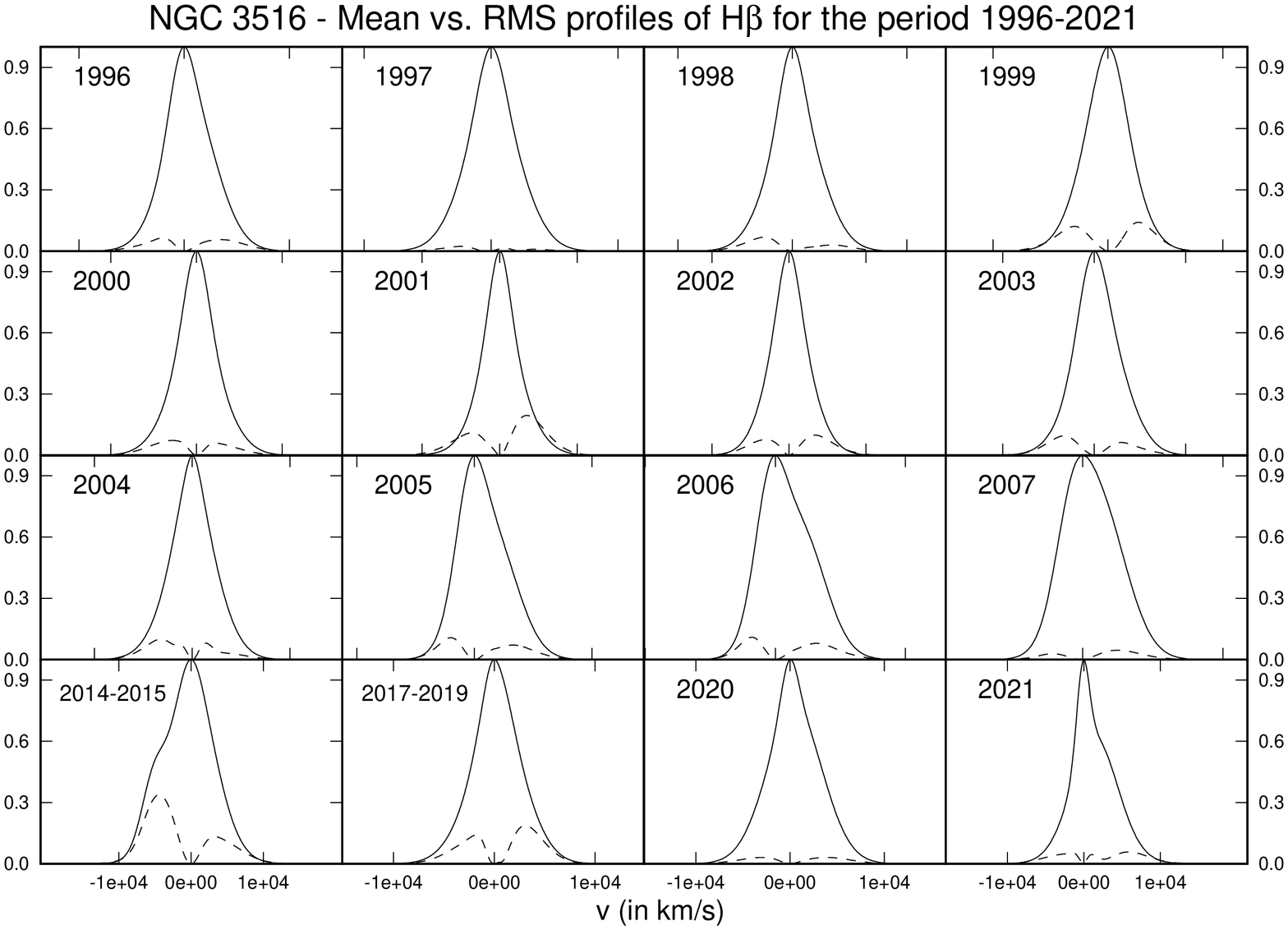}
\caption{Modeled broad H$\beta$ line shapes during the monitoring period. The  normalized (to unity) mean H$\beta$ profile (solid line) and corresponding rms (dashed line) in velocity scale is given for each year.}
\label{fig01a}
\end{figure*}

 In the period from 2020 to 2021, the broad component becomes stronger, but it seems  that there are strong changes in the line profile (see Figs. \ref{fig01} and \ref{fig01a}). This implies that the BLR geometry is changing in these two periods, after the BLR again appears. 

 \begin{figure*}[t!]
\centering
\includegraphics[width=0.7 \textwidth]{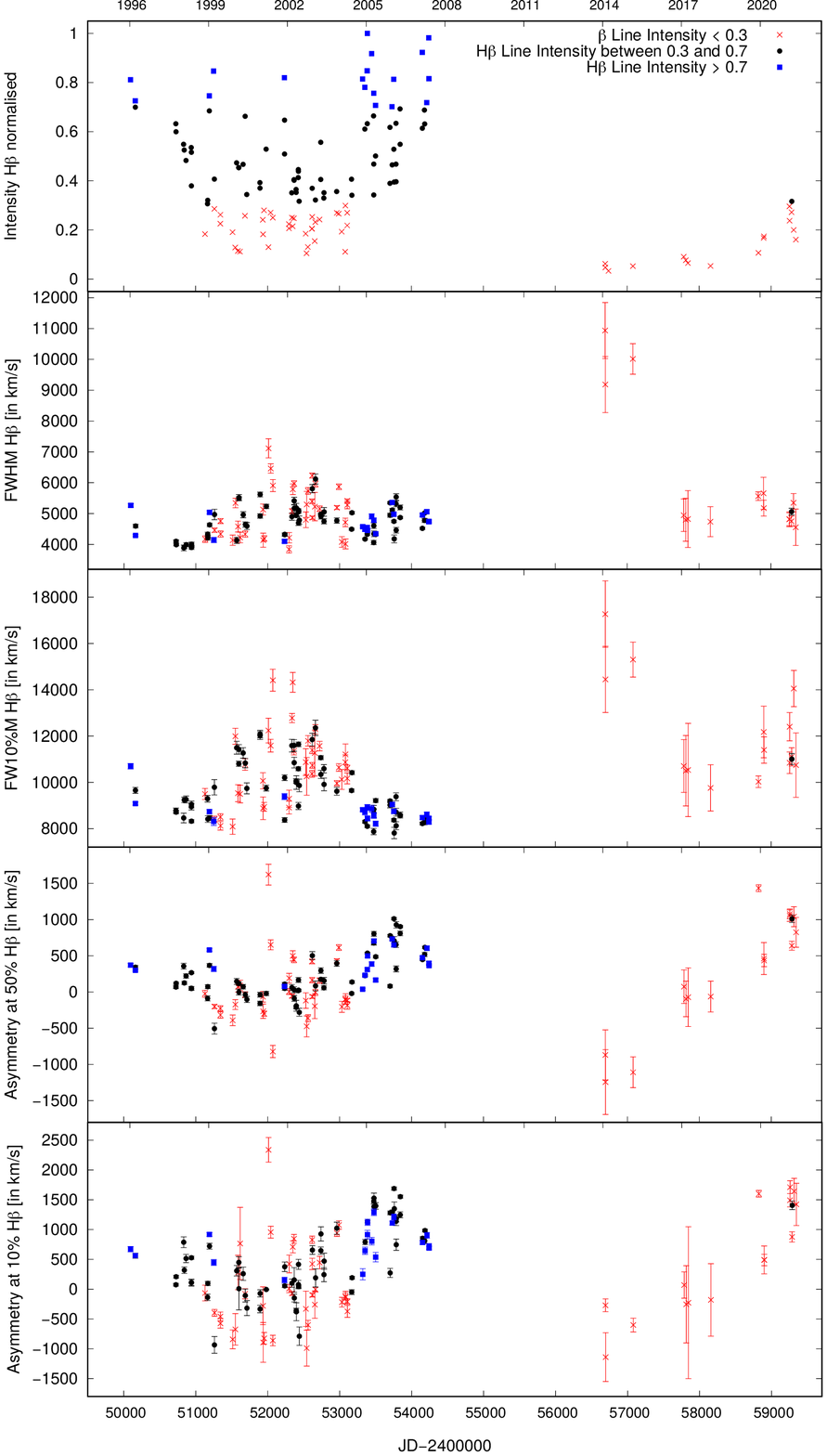}
\caption{Measured broad H$\beta$ line parameters during the monitoring period (from up to down): { Intensity (first panel),} FWHM ({ second} panel) and FW10\%M ({ third} panel), asymmetries measured at FWHM ({ forth} panel) and FW10\%M ({ fifth} panel). The used notation is: blue represents the maximum activity ($>0.7I_{max}$),  red the extreme minimum activity ($<0.3I_{max}$) and black represents observation between these two activity phase ($0.3I_{max}<I<0.7I_{max}$).}
\label{fig01b}
\end{figure*}

 \begin{figure*}[t!]
\centering
\includegraphics[width=0.94\textwidth]{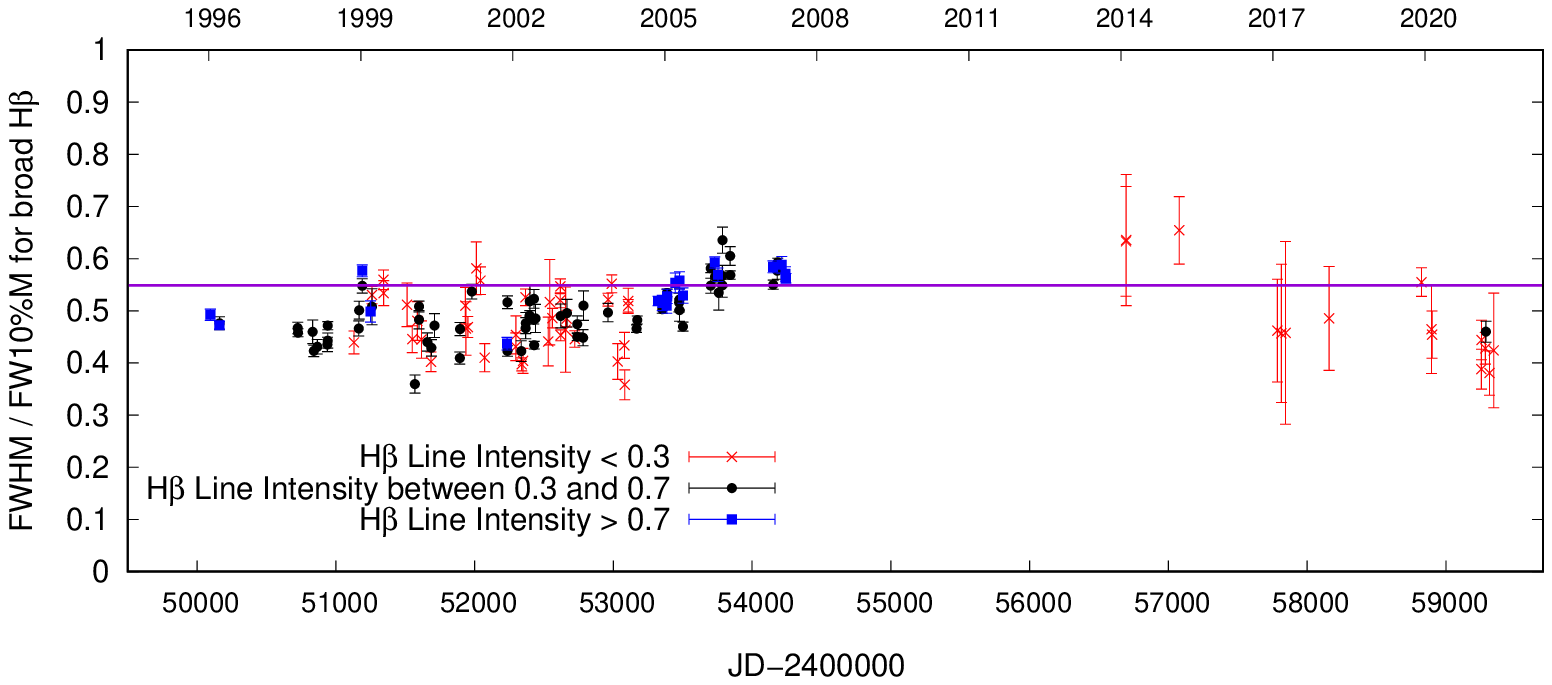}
\caption{Ratio of the FWHM and FW10\%M for H$\beta$ during the monitoring period. The horizontal line represents the value expected in the case of a Gaussian line profile.}
\label{fig01c}
\end{figure*}

To explore the changes in the line profile, we normalized  the line peak  (to unity) of the modeled broad H$\beta$ line and found the mean and  rms  profiles for each year when the line was observed (see Fig. \ref{fig01a}). As it can be seen in  
 Fig. \ref{fig01a}    the largest changes are during the phase of weak line emission. This can be caused by normalization, since the rms is normalized to the averaged profile, and in the phase of low line luminosity, the change is larger. However, the changes in the line profiles show larger asymmetry in the  low-luminous  line phase (2000 and 2001), as well as in the period 2014-2015; also a big difference between line profiles is present between 2020 and 2021. 

 We measured the H$\beta$ broad line parameters for each observational epoch. The broad H$\beta$ line for each epoch is modeled and normalized to one (as it is shown in Fig. \ref{fig01a} for averaged spectra) and after that we measured the line FWHM, Full Width at 10\% of Maximum (FW10\%M) and asymmetries at half and 10\% of the line maximum. The asymmetry is measured as the difference between the centers of the line at FWHM and at FW10\%M and the rest wavelength of H$\beta$.
 
 To explore the influence of different activity phases to the broad H$\beta$ parameters, we compared the line parameters from three activity phases  defined on basis to the  H$\beta$ line intensity taking the maximal intensity ($I_{max}$)  on Jan. 18, 2005 (see Fig. \ref{fig01}). We consider  the following three activity phases {(see first panel in Fig. \ref{fig01b})}: two extreme cases, the extremely weak line with $I<0.3 I_{max}$ (red points in Figs. \ref{fig01b} and \ref{fig01c}) and the phase with high line intensity $I>0.7 I_{max}$ (blue points), and the period of moderate activity (black points).
 
 In Fig. \ref{fig01b} we present  the line parameters variations during the monitoring period. The first two upper plots in Fig. \ref{fig01b} present the FWHM  and FW10\%M of H$\beta$ line.  As it can be seen in Fig.  \ref{fig01b} in the minimum of activity, the line width tends to be broader, while in the maximum of activity the width is narrower.  The measured FWHM is mostly around 4000-5000 km s$^{-1}$, while in the minimum, the width seems to be larger. The largest broad line measurements corresponding to the extreme minimum in 2014--2015 probably are affected by a large errors, but if we compare  the broad line intensity shown in Fig. \ref{fig01b} we can see that the broader (FWHM$>$5000km s$^{-1}$ and FW10\%M$>$10000 km s$^{-1}$) H$\beta$ line shows up mostly in the low line intensity phases (in the periods of 2001-2004 and 2014-2021 the observed H$\beta$ lines were weak).
 
To detect potential influence of inflows/outflows to the line profile, we measured the asymmetry of the H$\beta$ line at 50\% and 10\% of line maximal intensity taking the shift between the central peak\footnote{Estimated at 90\% maximal intensity in order to avoid the local line peaks caused by subtraction of the narrow H$\beta$ component} and line center at these values of maximal intensity.

The asymmetries at 50\% and 10\% of line maximal intensity are shown in Fig. \ref{fig01b} (third and fourth panels from top). As it can be seen,  there is an indication that in the minimal phase of line intensity a blue asymmetry is present, while in the maximum of the line intensity the  line profiles are more shifted to the red. It is interesting to notice that in the period of extreme minimum (2014--2021), the line shift is going from blue to red.  It may indicate that the BLR is changing geometry, which can be caused by several processes in the BLR, as e.g. Keplerian gas motion and outflows.

As we noted above, the line shape indicates the BLR geometry, and if there were strong changes in the BLR geometry they should reflect in the broad line shape. E.g. if the Keplerian motion is changed to an outflow, the ratio of the FWHM and FW10\%M should significantly change.  We explore the ratio of these two widths in Fig. \ref{fig01c}.  The ratio FWHM/FW10\%M stays between 0.36 and 0.69, and it is mostly smaller than the expected ratio in the case of a Gaussian line profile (middle horizontal line). It indicates a complex BLR structure, revealed by fitting the broad H$\beta$ line profile with two Gaussian functions.

\begin{figure*}[t!]
\centering
\includegraphics[width=6cm]{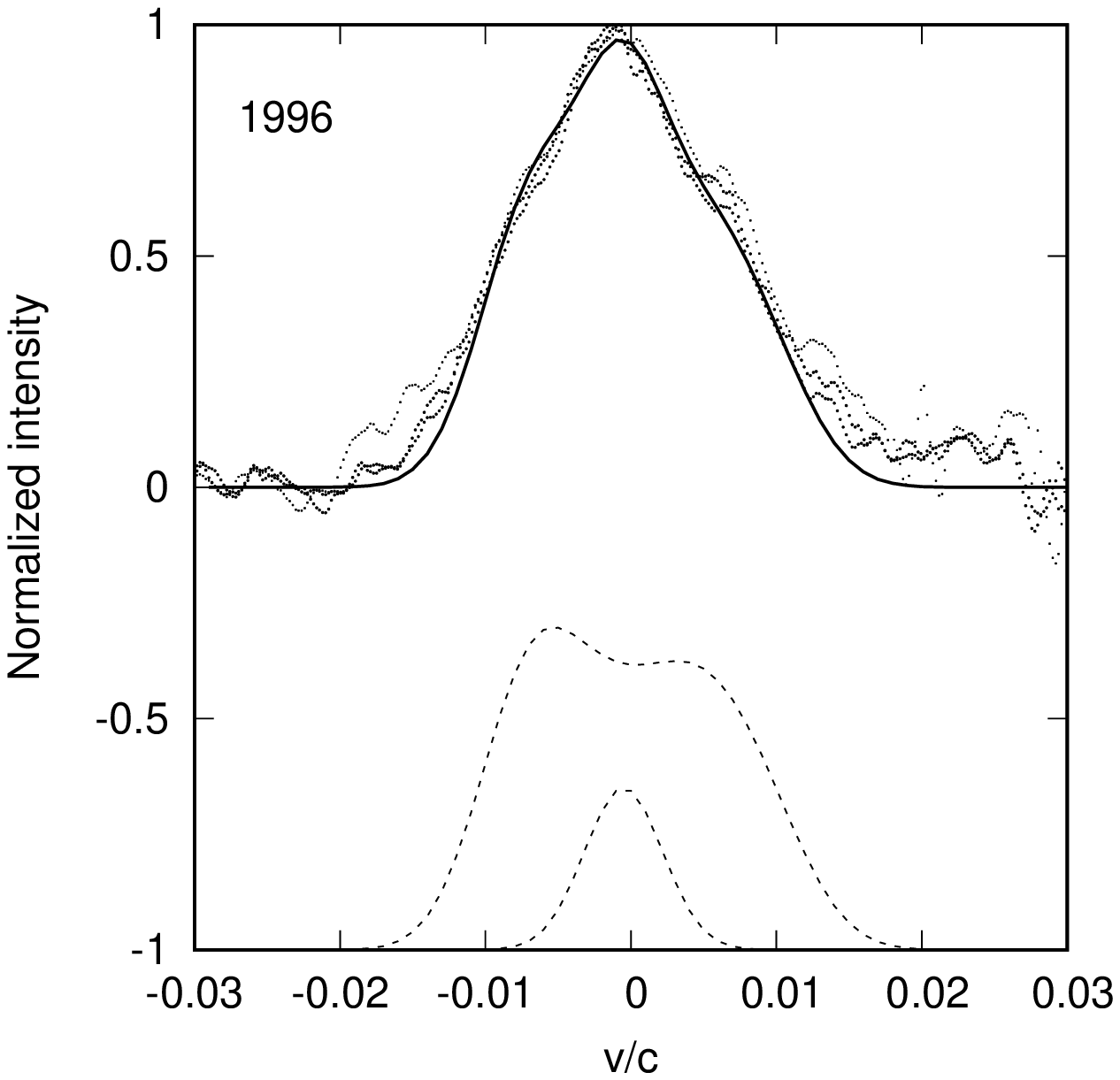}
\includegraphics[width=6cm]{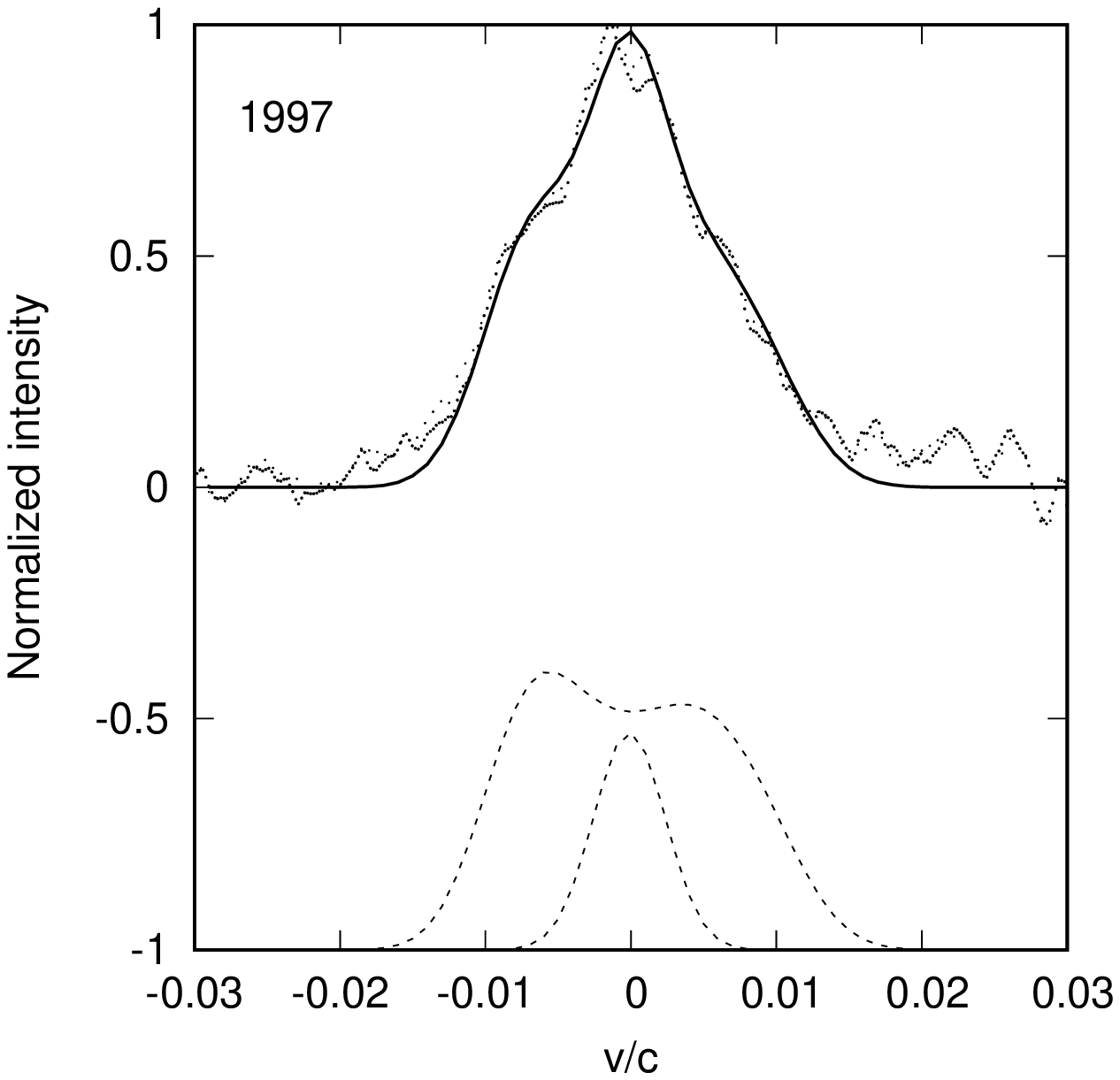}
\includegraphics[width=6cm]{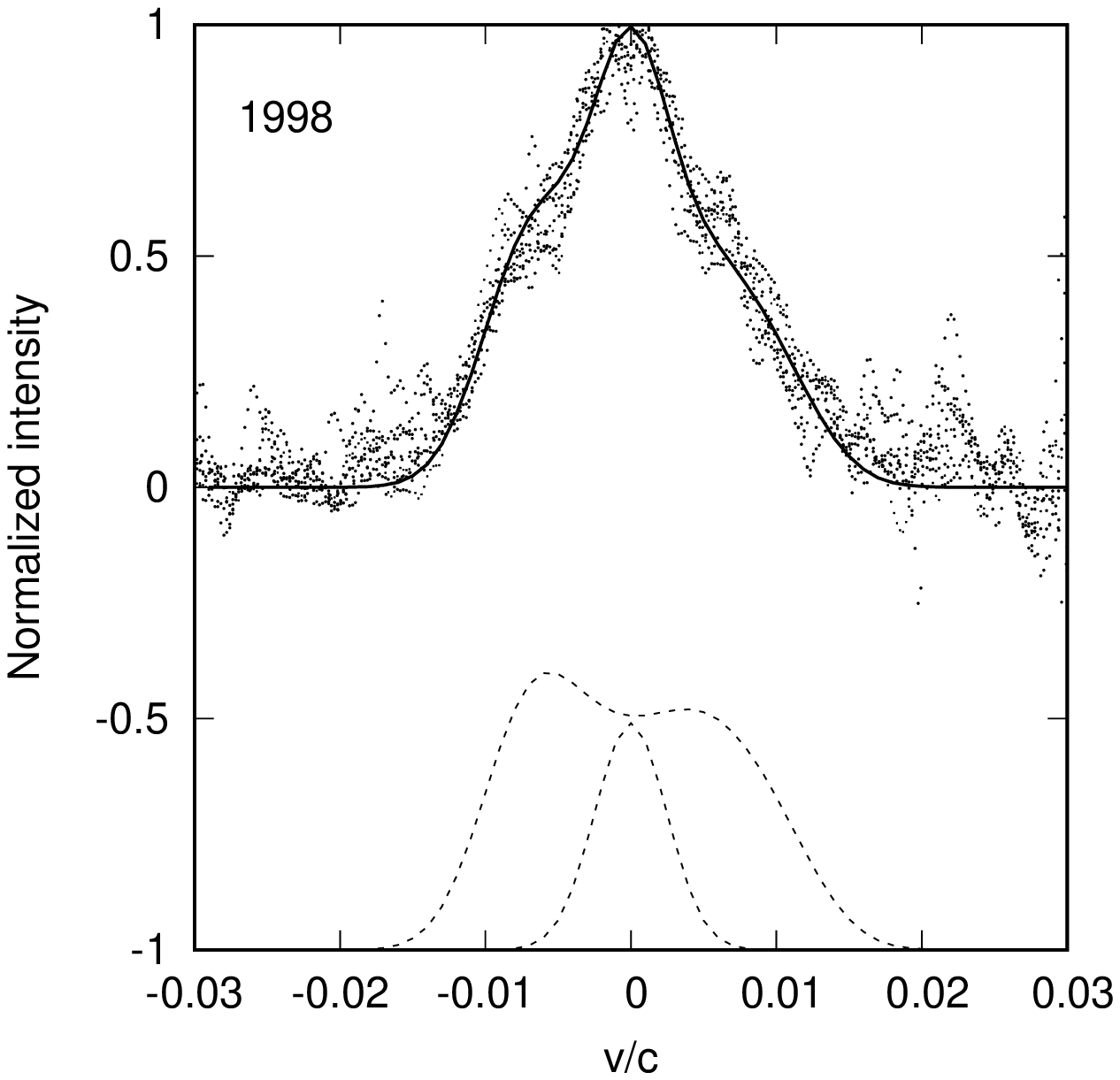}
\caption{Three observed H$\beta$ profiles in period 1996-1998 fitted with the two-component model (disc + one emitting region).}
\label{fig-mod1}
\end{figure*}

\subsection{Broad spectral line shape modeling: disc and an additional region}

The changes in ratio between the FWHM and FW10\%M indicate that the BLR is complex, therefore we consider a BLR with several components. We followed the investigation given in  \cite{po02}, where a two component model has been applied to the NGC 3516 line profile, showing that the emission line can be fitted with the emission from a disc like component (which mostly contributes to the line 
wings) and one intermediate component described with a Gaussian that represents an intermediate line region (ILR). There are a number of AGNs where the complex, two-component BLR model can describe the broad line profiles \citep[see e.g.][]{po04,bo09}. 
The disc-like BLR can be expected in a number of AGNs which show double peaked line profiles \citep[see e.g.][]{er94,po11,po14}. However the variability in the line profiles of double peaked lines can be
different from case to case \citep[see the case of two double peaked AGNs 3C390.3 and Arp 102B in][respectively]{po11,po14}.

\begin{figure*}[]
\centering
\includegraphics[width=6cm]{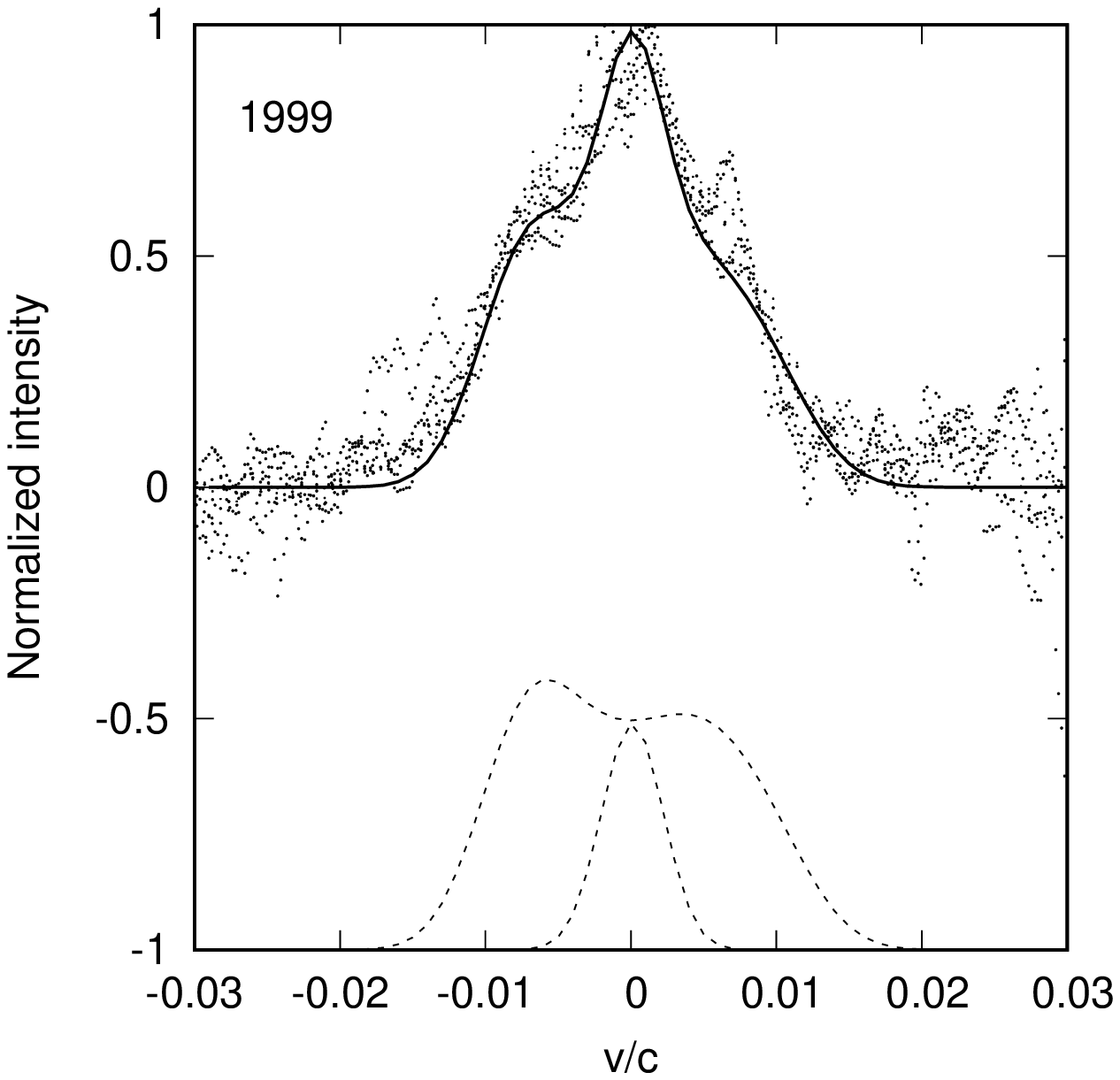}
\includegraphics[width=6cm]{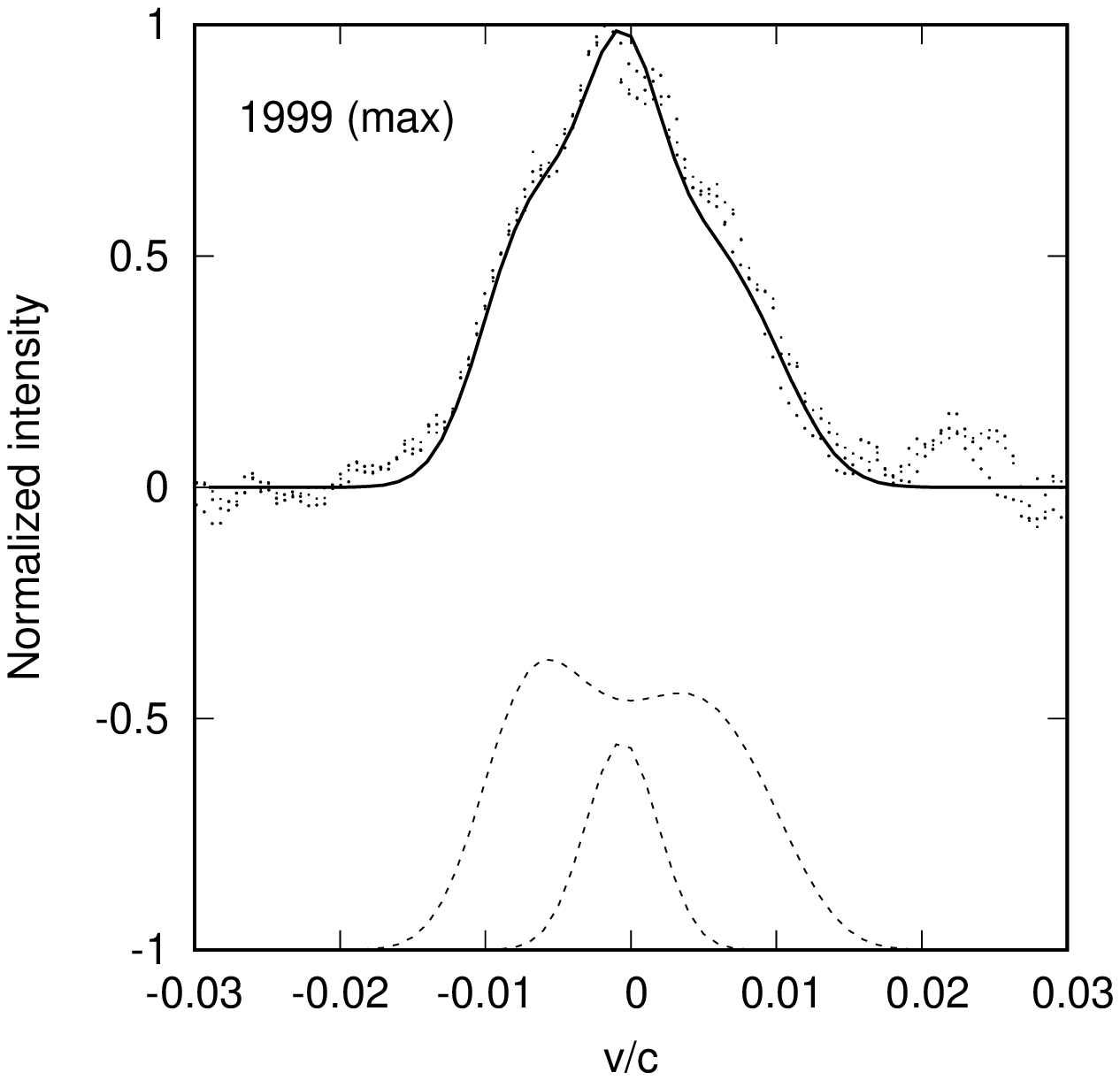}
\includegraphics[width=6cm]{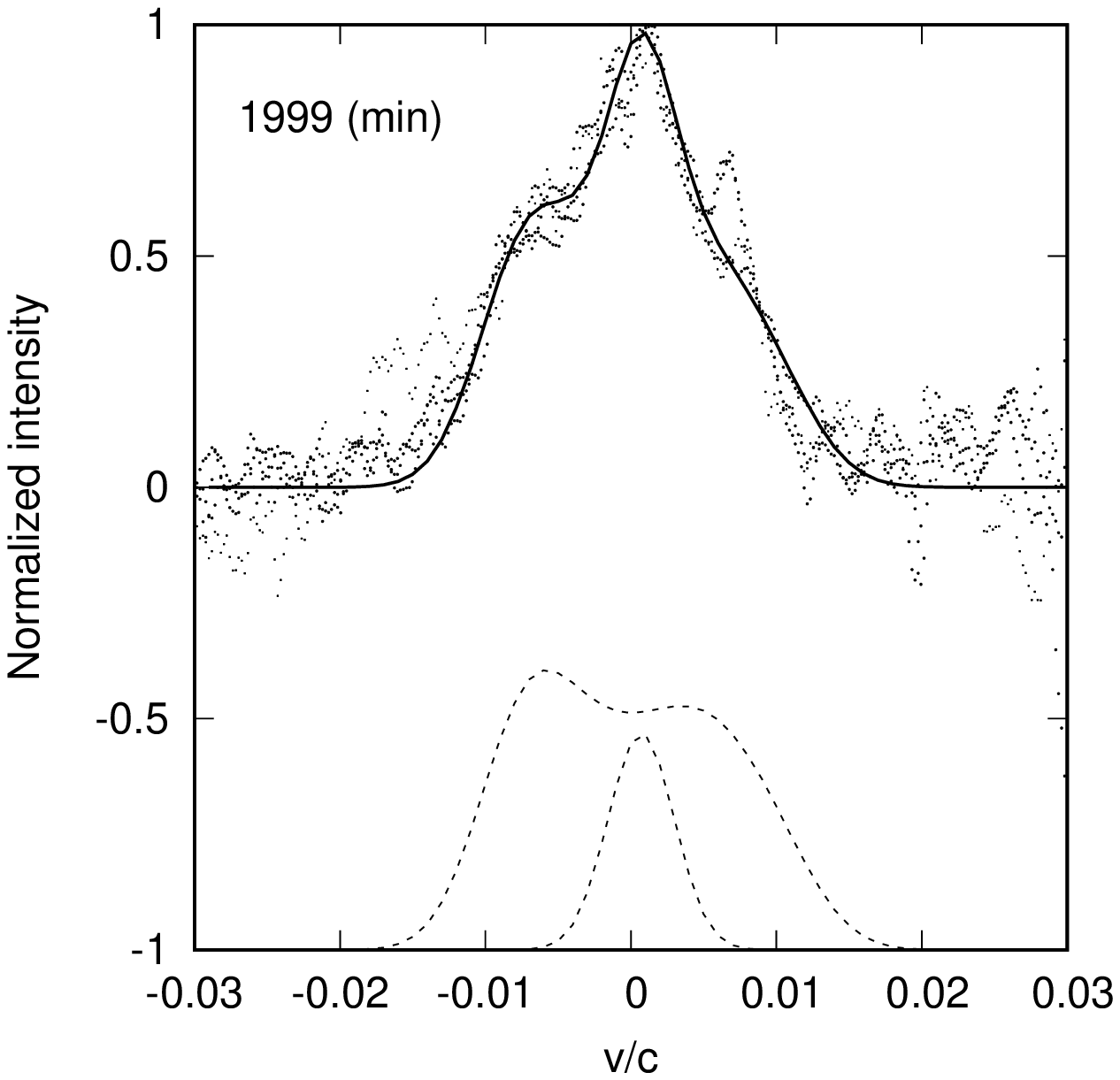}
\caption{The same as in Fig. \ref{fig-mod1} but for the averaged profile observed in 1999 (left) and in two activity periods: maximum (middle) and minimum (right).}
\label{fig-mod2}
\end{figure*}

 For the disc-model we used the model of \citet{ch89} and we added a Gaussian that represents the ILR emission\footnote{Note here the ILR + disc-like BLR decomposition is roughly equivalent to the BLR + VBLR (Very Broad Line Region) or ILR + VBLR decomposition considered by several authors \citep[see e.g.][]{fe90,su00,po04}}.   We found that the model can fit well the observed H$\beta$ profiles (see e.g. Figs.  \ref{fig-mod1} -- \ref{fig-mod2}).
   
It is hard to extract parameters from a best fit of this complex model, since the two-component model has a number of parameters that can be correlated. Therefore, here we fit the two component model to the average of all observed H$\beta$ lines during a year (see also Appendix A). We normalized the line profiles to one and converted wavelengths to velocity scales. As it can be seen in Fig. \ref{fig-mod1} the model can well fit the broad H$\beta$ profiles. In Fig.  \ref{fig-mod1}  we show  fits of the two component model using the observations from the beginning of the monitoring period (1996-1998). In some cases, where we have larger variability in the line intensity we divided the set of profiles between  high activity, and profiles with low activity as it is in the case of 1999 (see Fig. \ref{fig-mod2}).

\begin{figure*}[]
\centering
\includegraphics[width=6cm]{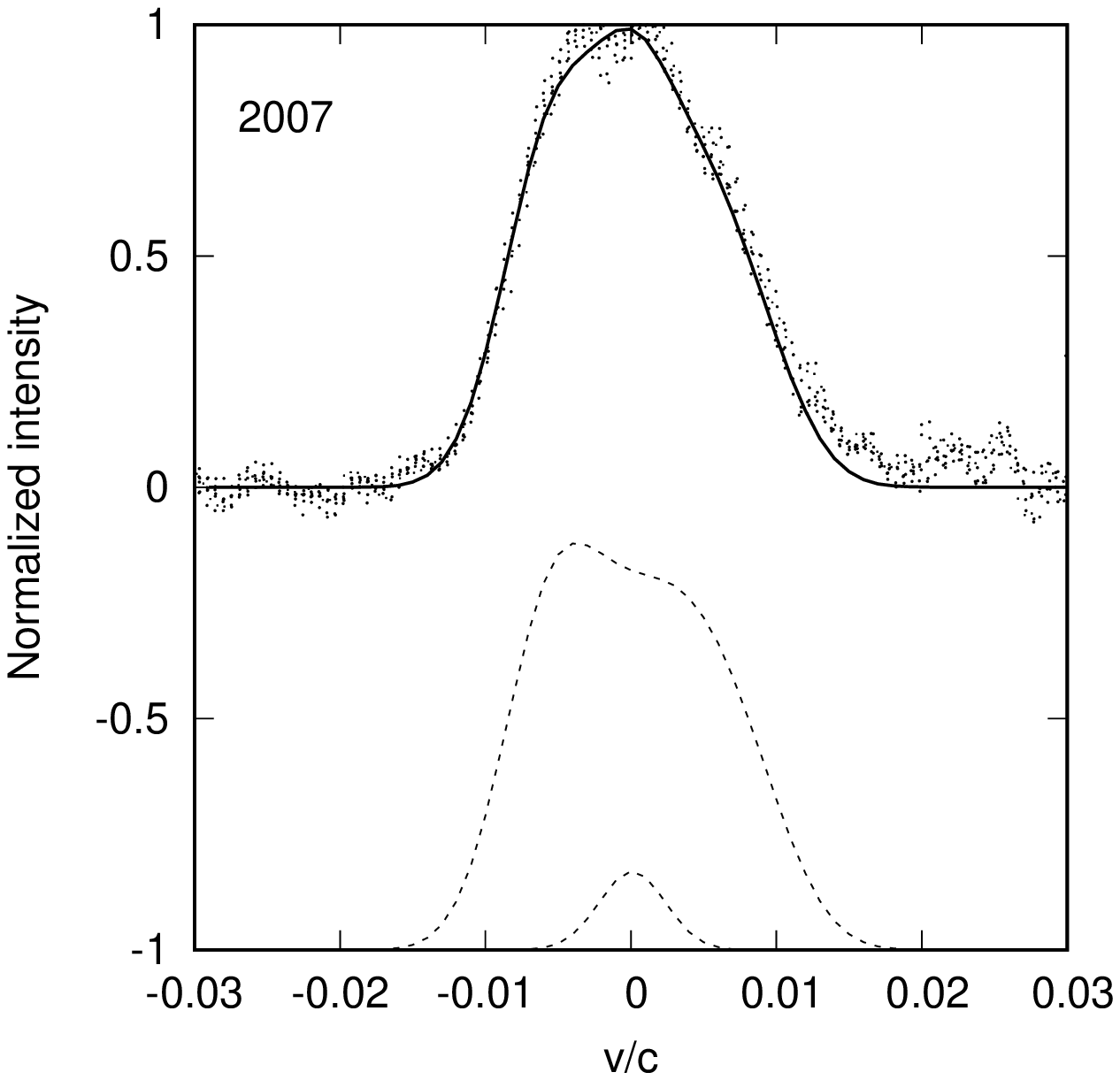}
\includegraphics[width=6cm]{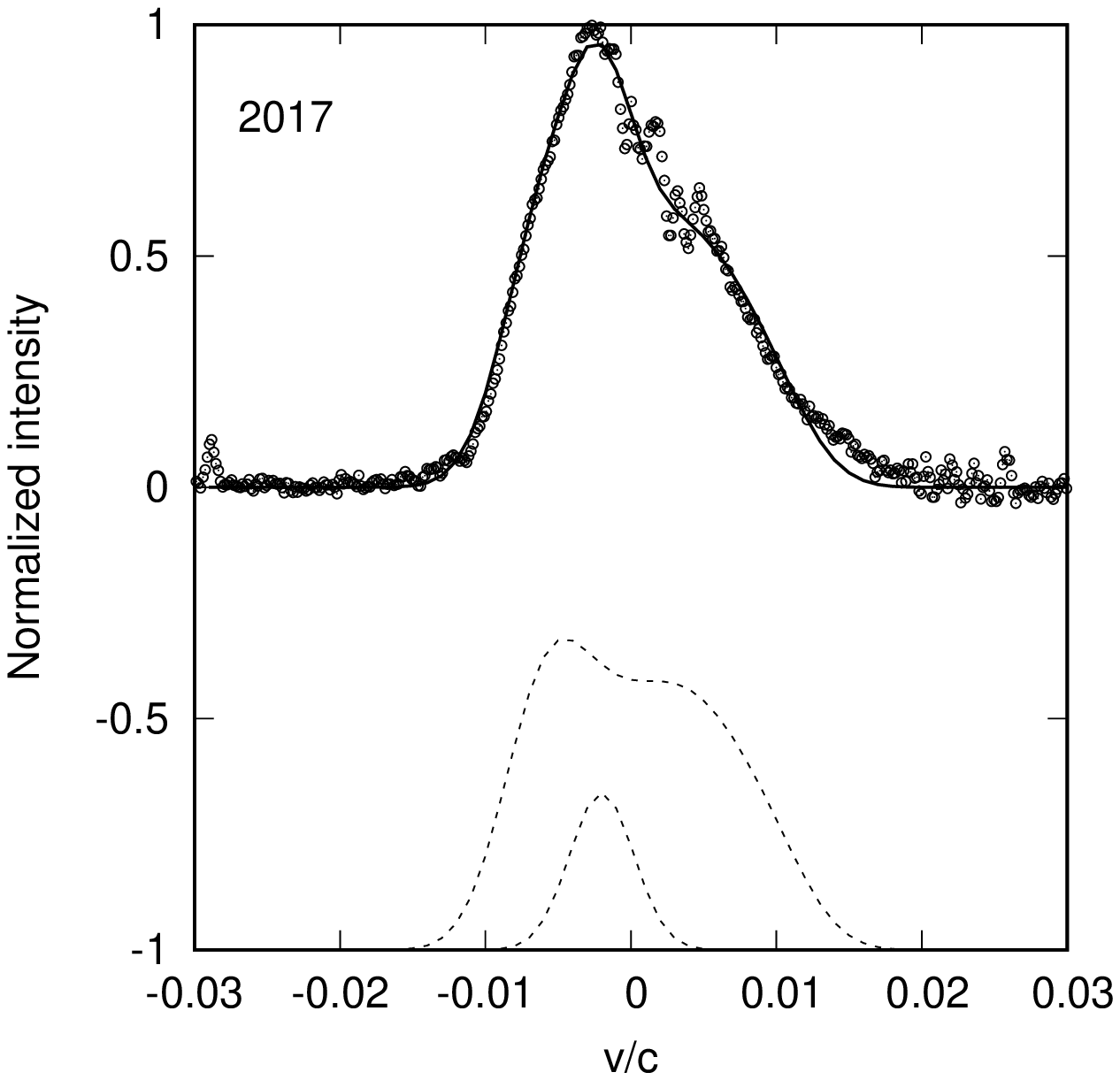}
\includegraphics[width=6cm]{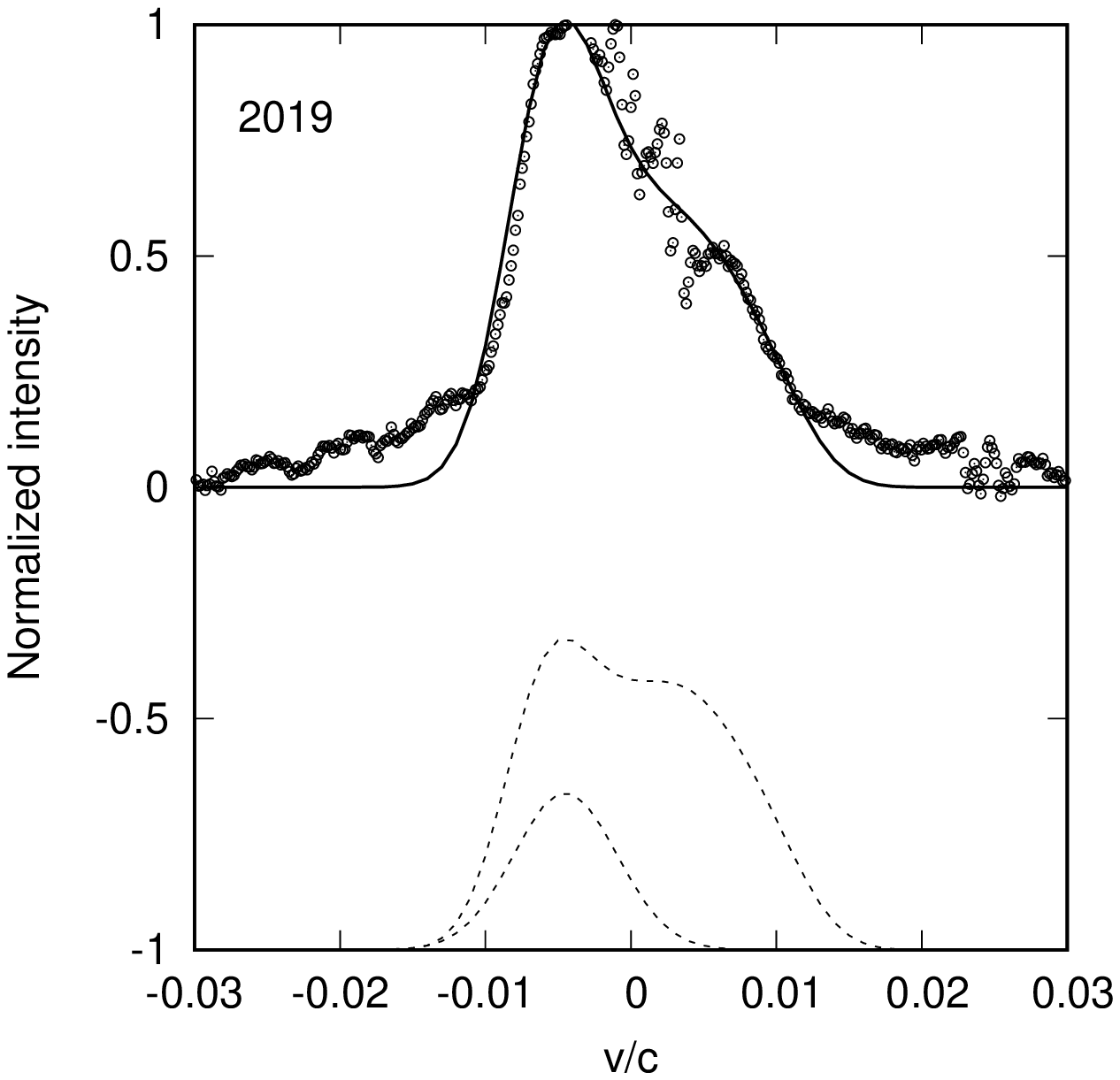}
\caption{The same as in Fig. \ref{fig-mod1} but for an averaged profile observed in 2006 and 2007, where line maximum intensity was observed.}
\label{fig-mod-max-min}
\end{figure*}

The line profiles from all periods can be, at least roughly, described with a two-component model, where  one component is Gaussian, mostly shifted to the blue and a disc-like component which is slightly shifted to the blue. There is also an indication that in the low activity phase, the central component is more shifted than in the active phase. As it can be seen in Fig. \ref{fig-mod-max-min}, comparing the high activity phase, which was in 2007, the central component is weak and is in the center of the line, while in the phase of low activity (2017 and 2019), the central component is significantly blue-shifted. From the fit we found that in 2007, the shift of the central component is close to zero, while in 2017 the shift is 
-600 km s$^{-1}$, and in 2019 is around -1500 km s$^{-1}$.

The obtained disc parameters
show that there is no big change in the disc parameters, only the intensity of the disc contribution to the line has been changing (as well the intensity ratio between the disc and the ILR component has been changing). We found roughly that the disc inclination is around 11-12 degrees, the inner radius around 320-420 gravitational radius, the coefficient of the disc emissivity\footnote{The emissivity is assumed as $\varepsilon\sim r^{-q}$ \citep[see e.g.][]{po02,po04}} is around $q\sim 3$, and the disc line is  blue shifted (around -600 km s$^{-1}$).  This may indicate the presence of a spreading  disc-like region \citep[see e.g.][similar as  the disc-like BLR obsereved in 3C390.3]{po11}.

\

\section{Discussion}

In order to find the nature of the change-look mechanism in the case of NGC 3516, here we explore the H$\beta$ line shape in a period of 25 years. The Type 1 AGN spectra of NGC 3516 changes  after 2014 to  a typical Type 2 spectra. Following the line shape changes we find that the BLR seems to be complex, as it is found in \citet{po02}, showing two-component emission: one coming from a very broad disc-like region, and another which has a narrower Gaussian-like emission. The changes in the H$\beta$ line shape may indicate the change in the BLR geometry.

The maximum of the broad H$\beta$ line occurs in the 1990s, and after 1999, intensity variability remained more prominent (see Fig.  \ref{fig01}). The extreme minimum in broad line intensity was observed after 2014. Later on, the broad lines began to appear again with low flux, and from 2017 to 2020, we were able to detect broad H$\beta$ and H$\alpha$ lines \citep[see][]{sh19,il20}. In 2021, a trend of increasing H$\beta$ intensity is visible.

The fundamental question we hope to answer with this paper is the mechanism of NGC 3516's changing appearance. As we noted above, the changing appearance of an AGN can be also caused by obscuring material (absorption). In such a case, the obscuring material could have a patchy distribution, then the dynamical movement of dust clouds can result in a change of the continuum (and broad line) emission, that affects the current  classification.

\subsection{Nature of CL mechanism in NGC 3516}

The nature of the CL mechanism in NGC 3516 has been investigated in several papers \citep[see e.g.][etc.]{sh19,il20,ok21,me22,rt22}.
We take into account the two most likely scenarios for the changing-look mechanism discussed in the literature  \citep[see also][and references therein]{rt22}. The first where CL is due to changes in line-of-sight obscuration, i.e., the dust covering the accretion disc and the BLR emission, and the second where CL is due to intrinsic variability, i.e., the emission of the accretion disc and BLR is disappearing.

The first scenario for NGC 3516 was suggested by \citet{ok21} on the basis of the correlation between the  UV/optical and X-ray variations. They found that the correlation was present in the first half of 2020, and after that correlation disappears, which may indicate an obscuration by Compton-thick clouds crossing the line of sight.

The second scenario, or intrinsic variability, is caused by some physical mechanisms at inside the central part of an AGN. This  can be investigated using X-ray radiation or the phenomenon of accretion-powered radiation. \citet{me22} compared the X-ray flux of NGC 3516 in   2006 when the H$\beta$ line was strong (see Fig. \ref{fig01}) and in the low  line flux intensity (observed 2017). They found that the intrinsic bolometric luminosity of NGC 3516 was lower by a factor of 4–8 in 2017 than it was in 2006. Contrary to \citet{ok21}, they suggested that that the CL mechanism is caused by the change of the ionizing spectral energy distribution (SED), without obscuring effect.
Furthermore, studies of the reverberation of the optical and Fe K$\alpha$ lines favored the intrinsic mechanism of CL AGN in NGC 3516 \citep[see][]{fe21,no22}. The variability in the broad line fluxes obtained by \cite{fe21} supports the intrinsic variability hypotesis. This has been confirmed by \cite{no22}, who investigated the variability of the Fe K$\alpha$ narrow line. Based on the narrow Fe K$\alpha$ line variability in the low-phase (2013--2014, when the broad H$\beta$ line disappears), they discovered that the iron line emitting region has a radius of 10 light-days, which is consistent with the BLR radius in the broad line bright phase. It is interesting that they conclude that the BLR material remained at the same locations as in the Type 1 phase, but there were some effects that extinguished the BLR emission.

The results in the line shape changes of NGC 3516 are in favor of the intrinsic variability to explain the low BLR emission.  It is difficult to explain the variability of the H$\beta$ line shape by the obscuring effect because of the following facts: a) in the minimum activity or phase of the weak broad lines, the line shapes indicate the presence of the disc-like region and an additional region contributing to the line core similar (almost the same) as in the maximum of activity. The totally or partly obscuration should affect the geometry covering partly or totally the disc emission and/or emission of the region which contributes to the line core, and consequently one can expect quite different broad line shapes in minimum and maximum of activity; b) the observed change in the line profile cannot be explained by crossing the dusty clouds across line of sight between an observer and the BLR, since in this case one can expect velocity dependent line shape changes that are not observed; c) after a total minimum (when broad lines almost disappeared), the weak broad line appears in a spreading disc-like BLR, indicating that this effect is connected with intrinsic radiation. 

We can also consider a scenario in which the small accretion rate and therefore smaller flux of ionizing photons almost extinguish the BLR emission caused by the smaller rate of recombination. In support of this idea, we found that the broad line profiles were complex in both phases, maximum and minimum. It indicates that the BLR geometry has changed by some intrinsic processes, but not by the obscuring material. Furthermore, we can see that the line is broader in the minimum H$\beta$ intensity, indicating that the majority of emission is coming from the part of the BLR closest to the central black hole. It indicates that the changing-look effect in NGC 3516 is caused by the accretion and lack of photons from the central continuum source, which can only ionize the surrounding material\footnote{In E1 context NGC 3516 belongs to spectral type B1, and can be well understood as a low-accretor in Population B or even extreme Population B, as outlined in \cite{ma22}. Therefore at very low accretion rate only the gas in the disc-like region is illuminated}. This scenario is also supported by the lack of  X-ray emission in this period. Any lack of accretion (which could be caused by a variety of factors) could result in the absence of the ionization continuum and, as a result, in
 extremely weak broad emission lines \citep[see][]{ki18,no18}. Mrk 1018, for example, changed from Type 1.9 to 1 and back to 1.9 over a 40-year period, and it appears that NGC 3516 has made Type changes in the past, as the H$\beta$ show a broad line in the first observations performed by Seyfert  \citep[][]{se43}, and a narrow line in 1967, \citep[see][]{an68}.

Comparing the CL behavior of NGC 3516 with other AGNs, we see that  there is something common with Mrk 1018, since both AGN in the phase of Type 1 showed complex broad Balmer lines 
which indicate more than one emission line region \citep[][]{po02,ki18}. It seems that we have a disc-like BLR with an ILR which is emitting a Gaussian-like component, that is mostly shifted to the blue, indicating some kind of outflow in the BLR. We cannot exclude that the outflowing component may be associated with the innermost part of the Narrow Line Region, as it was noted in 
\cite{ma22}. However, a strong absorption in the broad UV lines \citep[][]{go99}, as well as in X-ray \citep[see][etc.]{kr02,tu11,ho12} is in favor of the BLR outflowing part.  The high-ionization emission lines 
(Ly$\alpha$ $\lambda$1215, C IV $\lambda$1549, N V $\lambda$1240, and He II $\lambda$1640) showed significant variation that was order of a factor of $\sim$ 2 \citep[see][]{go99}, similar as it was
found in the H$\beta$ and H$\alpha$ line variations \citep[][]{sh19}.

The low X-ray  emission was also observed in the period 2013--2014 \citep[][]{no16}, and was at level of just 5\% of the averaged X-ray flux from the  1997--2002 period \citep[][]{no16}. However,  new results obtained from the narrow Fe K$\alpha$ line showed that the BLR material is present in the phase of low broad line emission, and there are probably some mechanisms that can affect the rate of recombination and consequently turn off the BLR light. We found that line profiles can be described by the emission of the disc-like BLR and an additional emission that contributes mostly to the core of the H$\beta$ line. Also, the disc-like component were present in the low activity phase, which is in agreement with the finding of \citet{no22} that the BLR material is present in the Type 2 phase, but the effect of recombination cannot produce strong broad lines. 

To clarify the nature of the CL AGN in NGC 3516, we should  take into account the previous discussion that can be summarized as following:

\begin{itemize}
   \item Because broad H$\beta$ line shape variation, as well as X-ray or Fe K$\alpha$ line variation are detected, the obscuring effect is ruled out. In addition, it is hard to explain how obscuration can change the line profiles while  the broad line profile can be detected during most of the  monitoring period. It seems that there is a change in the contributions of different regions  to the total broad line profile.
    \item The BLR material remains the same  in both phases \citep[high and low broad line intensity, see][]{no22}, but the the change affects the rate of recombination, which contributes to the line intensity.  An additional problem is the intensity of the central continuum, since during the minimum the optical continuum is too low and the low continuum simply does not reach the gas further out.
    \item The disc-like geometry seems to be present in all phases during the monitoring period (except in total minimum 2014/2015). This confirms a rotational motion of emission gas in the BLR. This can be inferred from the broad line parameters. Also in the low line intensity phase, the line width seems to be broader indicating that most of the emission is coming closer to the central black hole.
\end{itemize}

\subsection{The role of dust and accretion in the CL mechanism}

To better understand the CL nature of NGC 3516, let us consider the spectro-polarimetric observations of a number of AGNs, where equatorial scattering in the inner part of the dusty torus is expected  \citep[see e.g.][]{sm95,po22}. The reverberation in the polarized broad lines, on the other hand, indicates that the scattering equatorial region is most likely coincident with the dust sublimation region, which is smaller than the estimated inner dusty torus radius \citep[see, for example, for Mrk 6 in][]{sh20}. This may indicate that some amount of dust is present in the BLR \citep[see][]{ga09,cz22} and the amount of dust can influence the rate of recombination in the BLR and consequently the intensity of the broad lines.

In the case of a dusty BLR, the scenario may be as follows. When the accretion rate is smaller, dust is  moving closer to the central source, making a dusty BLR. Since the ionization rate coefficients decrease when dust particles are  present in plasma \citep[see e.g.][]{li20} the rate of the BLR emission in the broad lines should become weaker, and the central continuum emission can be obscured, since dust may also extend in the polar region \citep[see e.g.][]{st19}. As the material is coming to the central black hole, the rate of accretion is increasing, and the rate of ionization is higher, so the amount of dust is smaller and the BLR emission is stronger, as well as the emission of the central continuum. In this scenario, the BLR material is also present during the lack of strong broad lines  as it was found in \citet{no22}. Additionally, in this case, the geometry of the disc-like region will be present all the time, and the blue-shifted central component (as well as the disc-like line) can be explained by the wind caused by radiation pressure.

Note here, that \cite{te22} suggested that in the majority of 412 Swift-BAT detected CL-AGNs, the CL mechanism is not due to changes in line-of-sight obscuration. Also,  changing in hard X-ray and optical suggests that the accretion drives the changes  between the Type 1 or Type 2. As obscuration by the dust outside the BLR would probably not explain the changes in the hard X-ray, the scenario of a dusty BLR can be considered as a model which can explain CL mechanism in a number of AGNs. Also, recent investigations of the nature of the CL in IRAS 23226-3843 show that it is  most probably caused by changes in the accretion rate \citep[see][]{ko22}.

\section{Conclusions}

We investigated the H$\beta$ broad line shape and  several related parameters emitted from the BLR of the AGN NGC 3516 observed in a 25-year period.
Based on our analysis of the changing broad H$\beta$  line profiles and parameters, we can outline the following conclusions:

\begin{itemize}

  \item The broad H$\beta$ line profile has changed during the period, indicating important changes in the BLR structure. The BLR structure seems complex.

   \item  The changes in the broad H$\beta$ line parameters indicate that the line seems to be broader in the low line intensity phase, which may be due to the emission of material that is coming close  to the central black hole. This indicates that the low intensity is probably caused by the rate of recombination in the BLR (or the rate of photoionization). This may be caused by a lack of ionizing photons, which can be caused by a lower rate of accretion or the presence of dust in the BLR.

  \item  The H$\beta$ line profiles can be roughly explained with a two component model, where one component represents a disc-like region, and another one is an intermediate component, which is shifted to the blue, indicating a wind from the center. The disc-like component is also shifted to the blue indicating a windy disc-like region.  Both regions are present in both cases: in the maximal and minimal broad line intensity. This also indicates that the emission from these two regions is disappearing due to some intrinsic effect in the BLR.
\end{itemize}

Finally, we can conclude that the change from Type 1 to Type 2 of NGC 3516 is most likely caused by  intrinsic effects in the BLR. This may be due to the lack of X-ray photons that can photoionize the BLR gas combined with dust plunging into the BLR. The photoionization in a dusty BLR has a smaller efficiency, and consequently, the rate of recombination is too small. This may  explain that the broad lines are almost disappearing from the spectrum.

%{\bf OVDE SI DOSTA ZAKOMPLIKOVAO, HAJDE DA SE CUJEMO I ZAJEDNO NAPISEMO} NADAM SE DA JE SADA JASNIJE

\section{Acknowledgments}

This work is devoted to our dear colleague Alla Ivanovna Shapovalova, who passed away in January 2019. She worked  despite difficulties until the very end on the first published  paper \citep[][]{sh19} as a pioneer and a leader of reverberation mapping campaign of AGNs at SAO RAS. This work is supported by the Ministry of Science, Technological Development, and Innovation of R. Serbia through projects of Astronomical Observatory Belgrade (contract 451-03-47/2023-01/200002) and the University of Belgrade - Faculty of Mathematics (contract 451-03-68/2022-14/200104).
 Observations with the SAO RAS telescopes are supported by the Ministry of Science and Higher Education of the Russian Federation. The renovation of telescope equipment is currently provided within the national project "Science and Universities". The work provided by A.B. and E.S. was performed as part of the SAO RAS government contract approved by the Ministry of Science and Higher Education of the Russian Federation.
L.\v C.P. and A.B.K. acknowledge the support by Chinese Academy of Sciences President’s International Fellowship Initiative (PIFI) for visiting scientist. D.I. acknowledges the support of the Alexander von Humboldt Foundation. 


\appendix

\section{Modeled line profiles}

Here we give the 
decomposition using a two-component model of the H$\beta$ broad lines profiles for different epoch. It is important that almost in all periods (excluding the minimum H$\beta$ intensity in 2014--2015), the line can be reasonably well fitted with a disc and an outflow component. 

\begin{figure*}[]
\centering
\includegraphics[width=6cm]{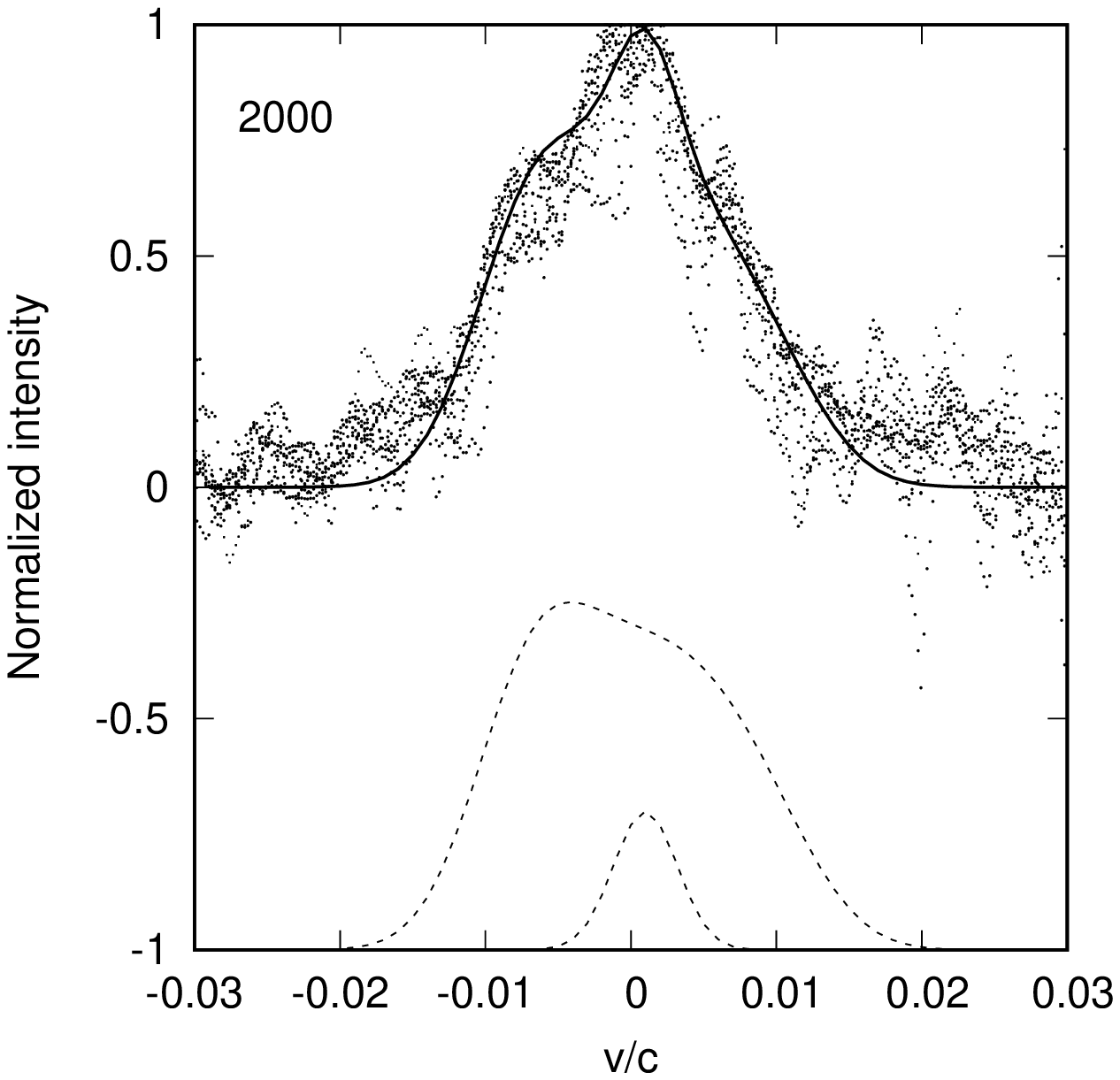}
\includegraphics[width=6cm]{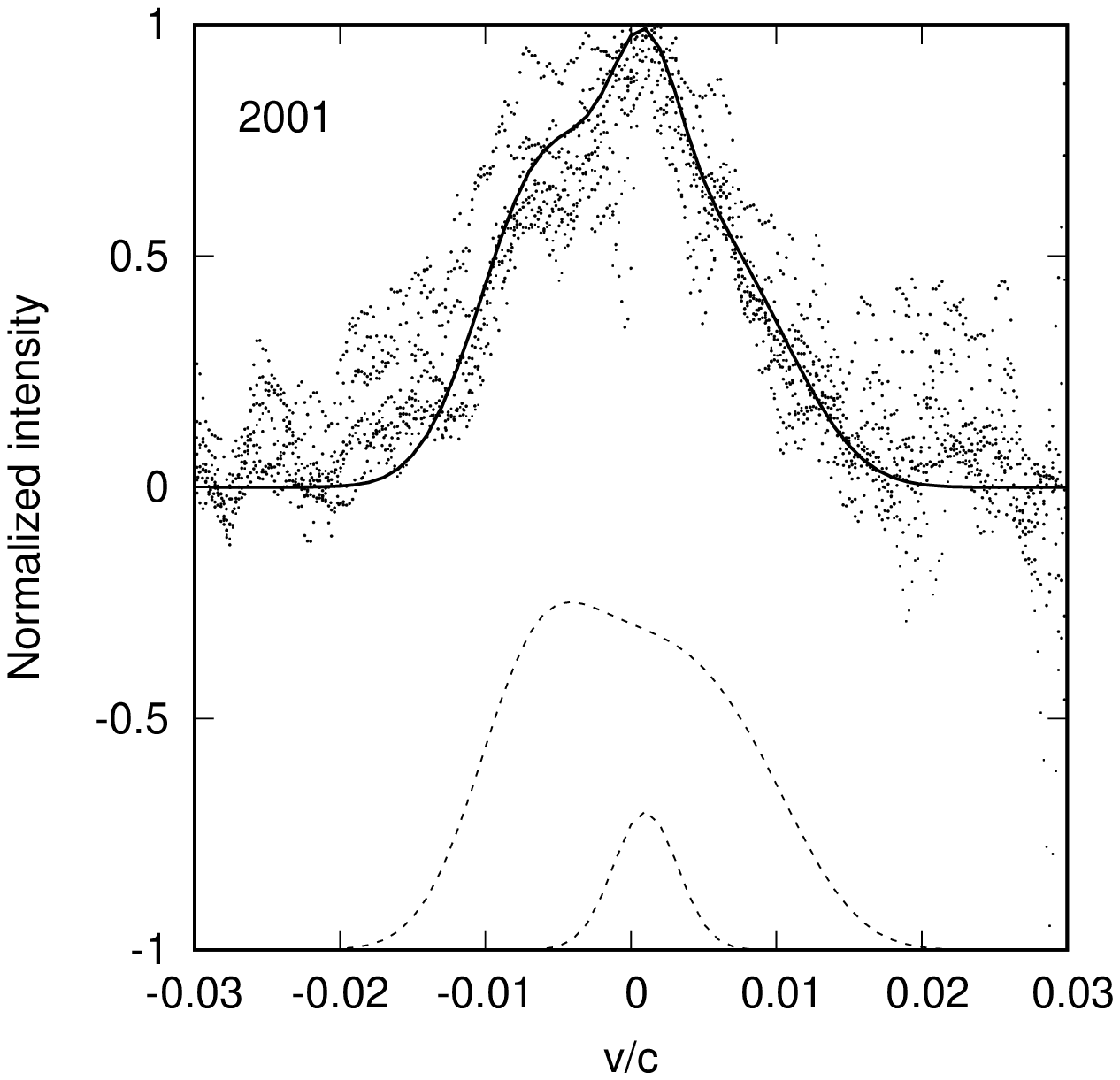}
\includegraphics[width=6cm]{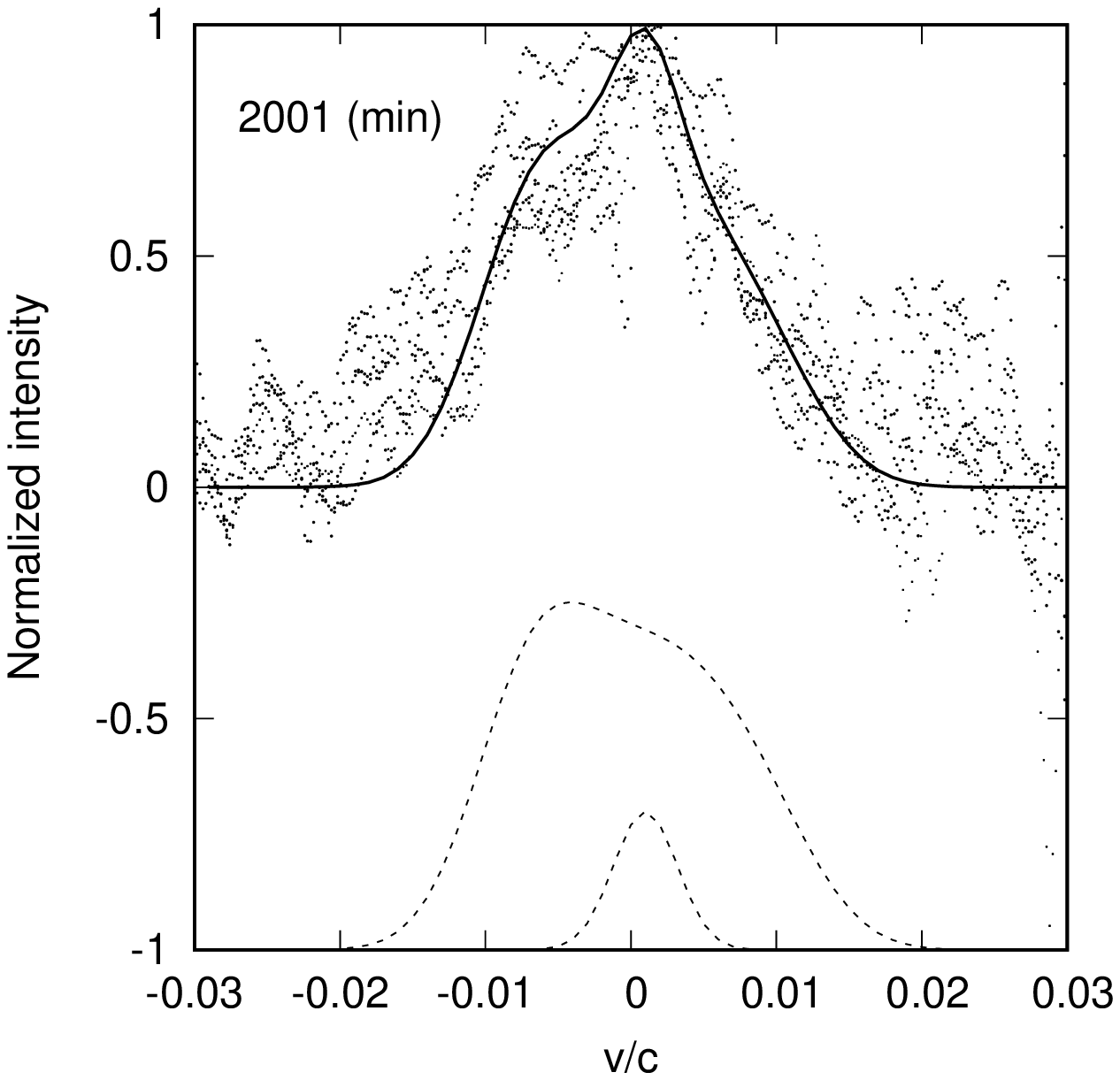}
\includegraphics[width=6cm]{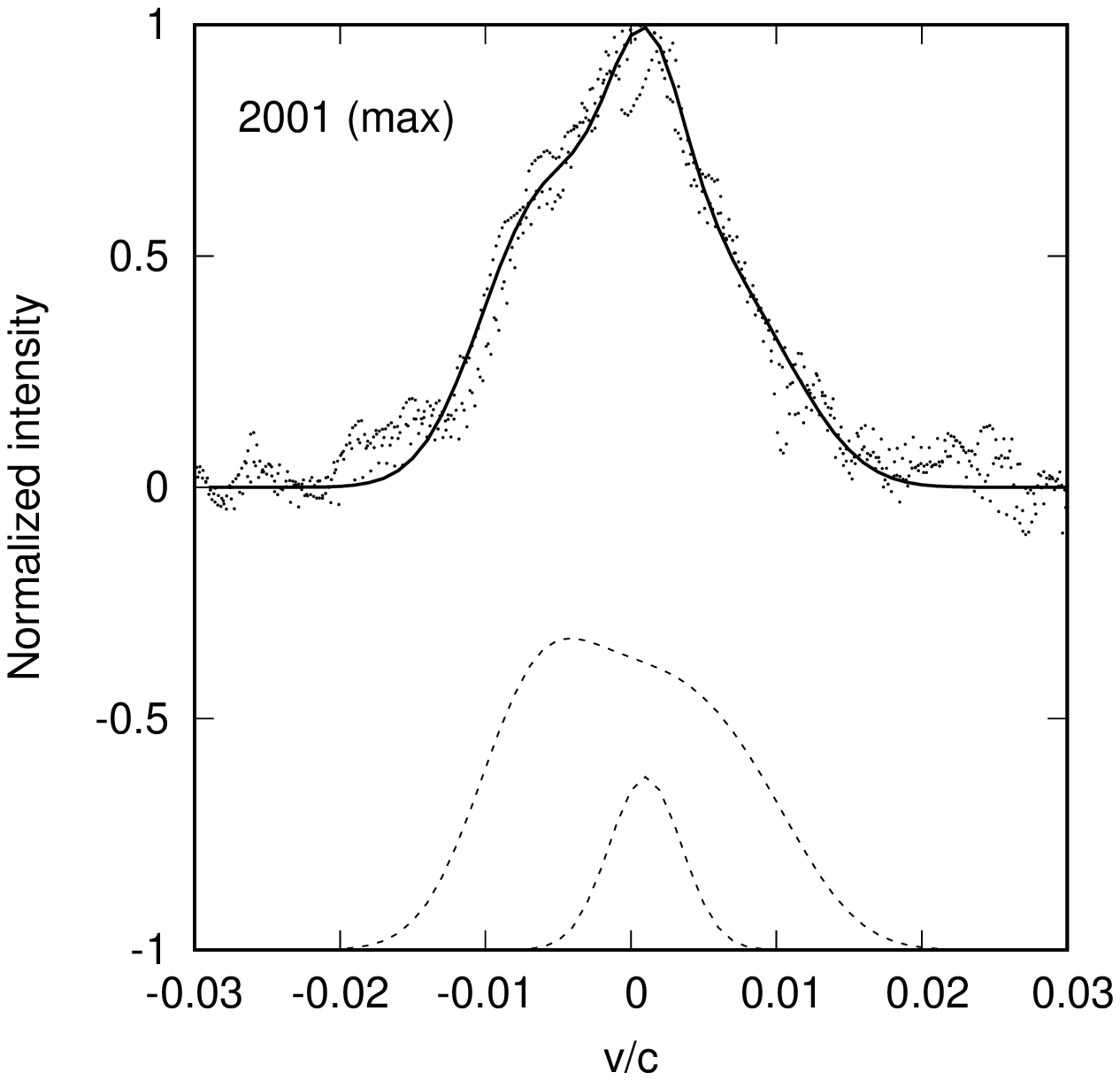}
\includegraphics[width=6cm]{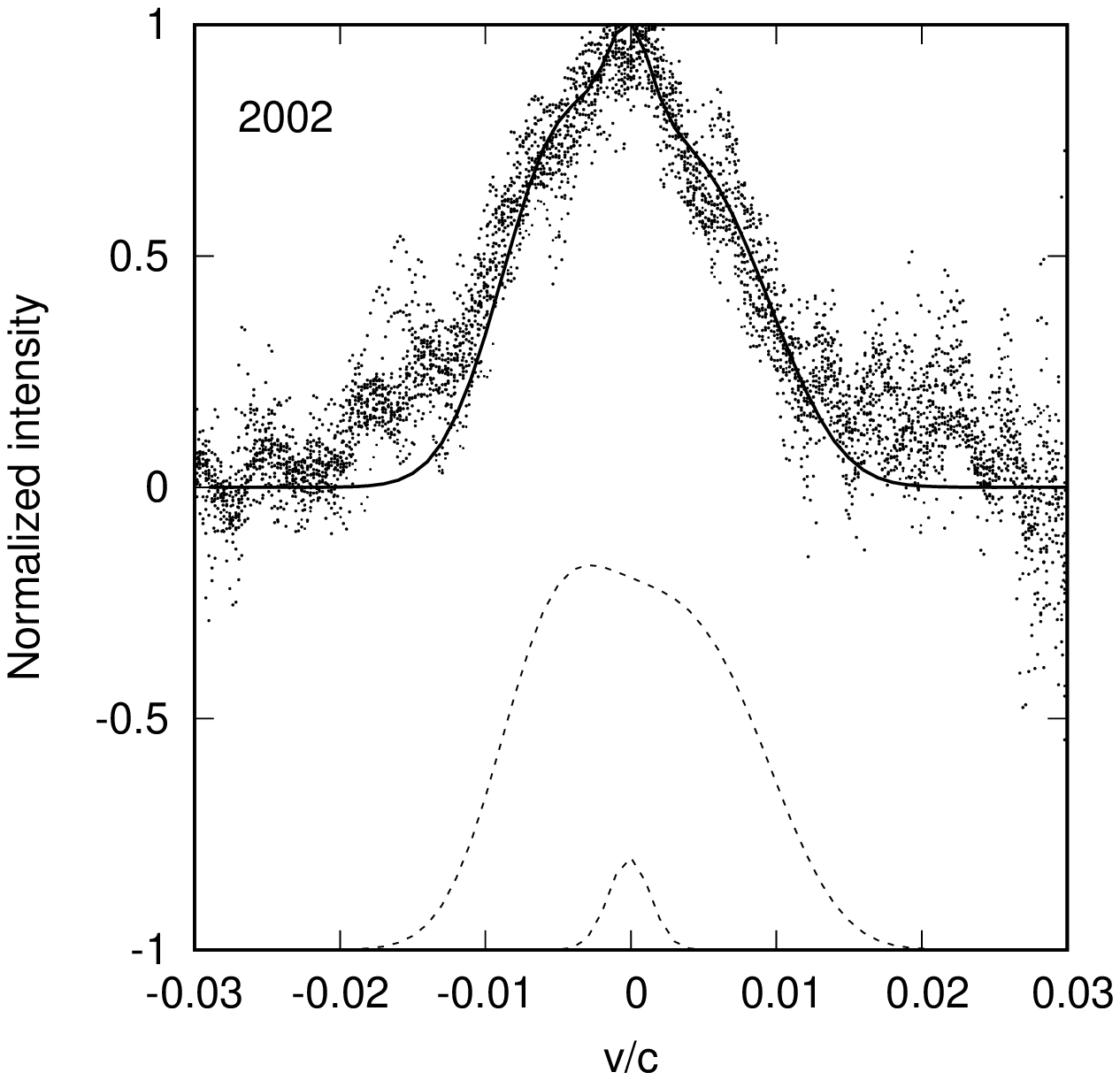}
\includegraphics[width=6cm]{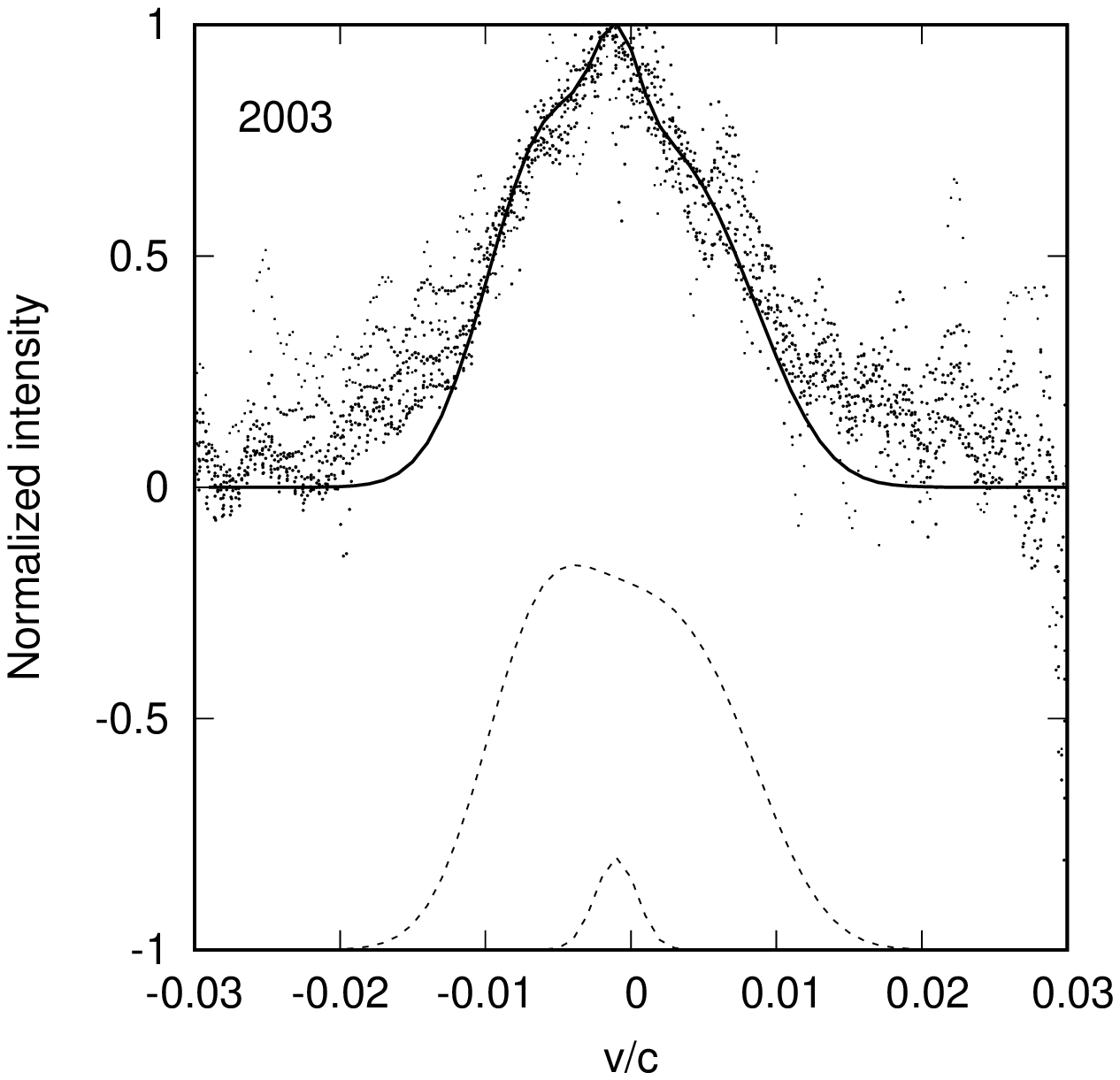}
\caption{The same as in Fig. \ref{fig-mod1} but for  averaged profiles corresponding to 2000, 2001 (whole period and minimum-maximum periods), 2002 and 2003, where the line was in the minimum.}
\label{fig-mod3}
\end{figure*}

\begin{figure*}[]
\centering
\includegraphics[width=7cm]{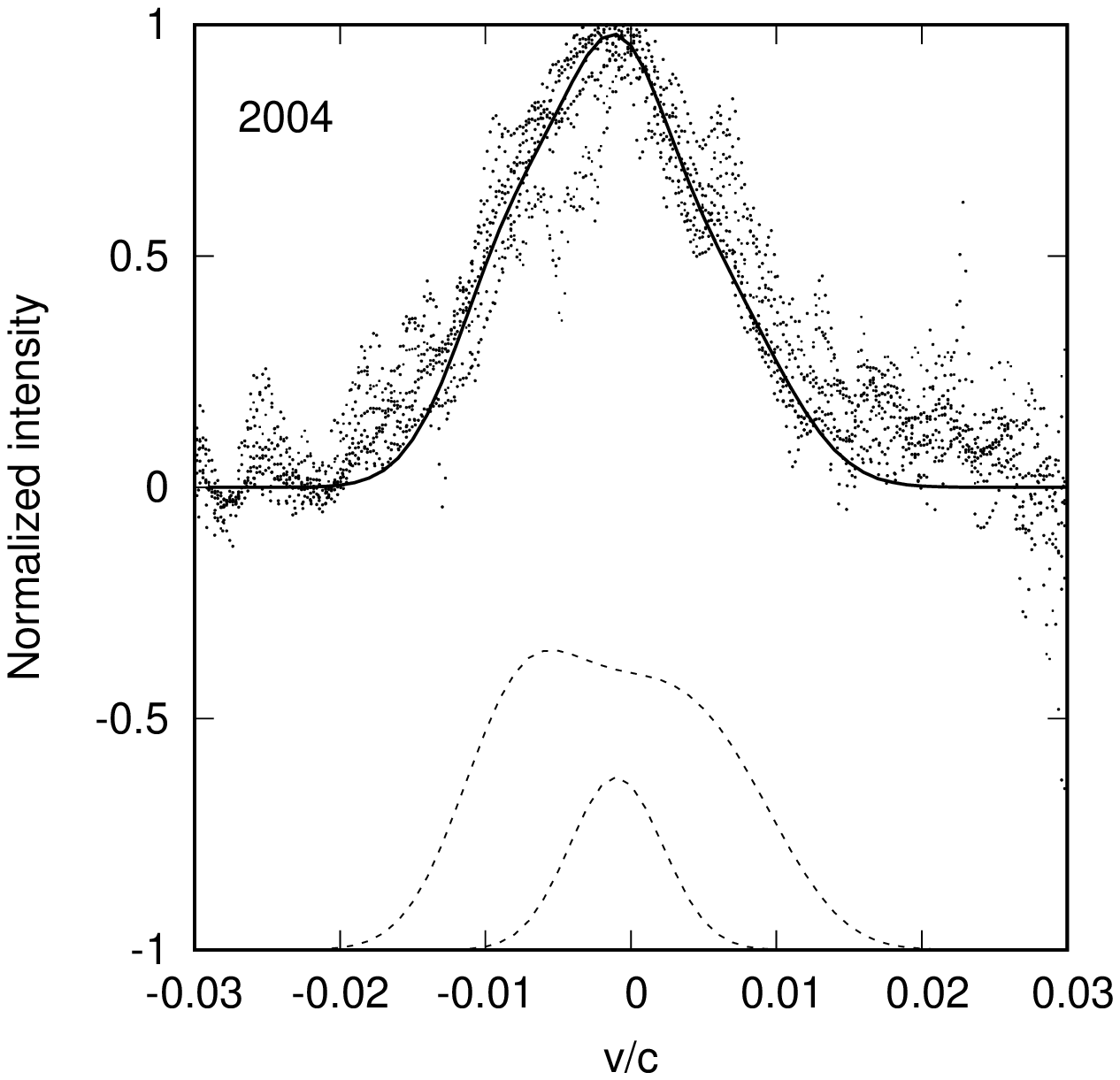}
\includegraphics[width=7cm]{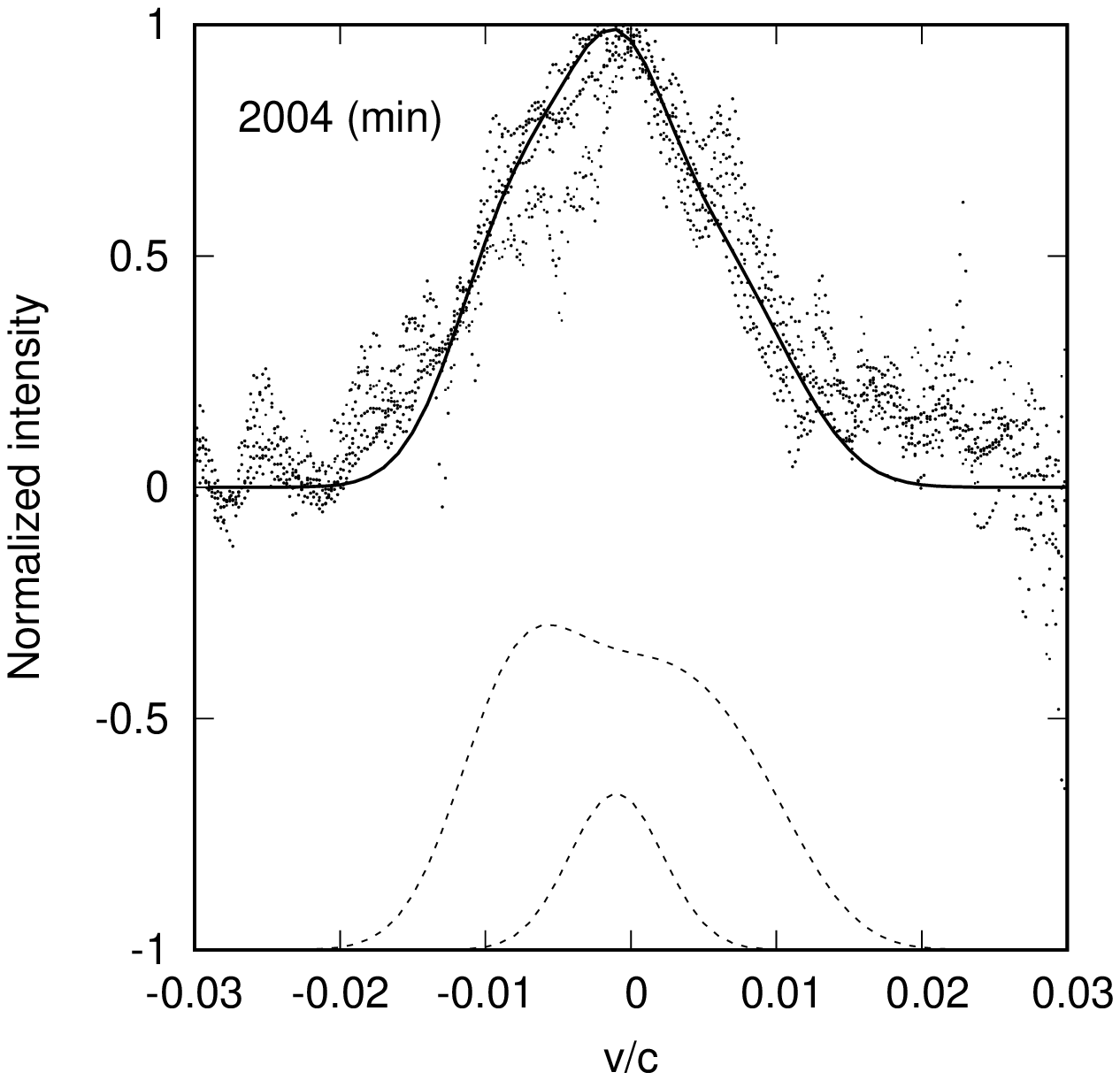}
\includegraphics[width=7cm]{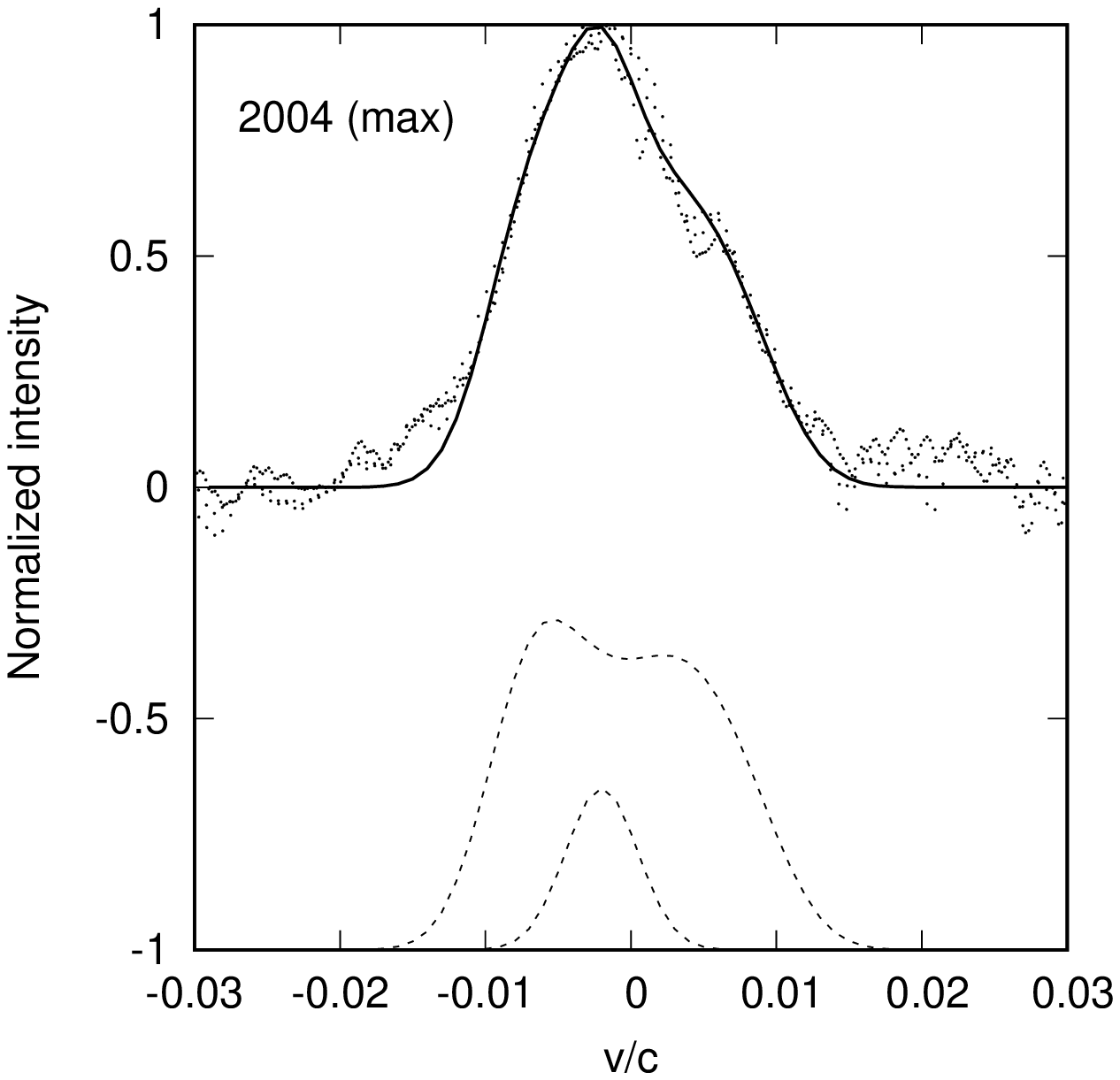}
\includegraphics[width=7cm]{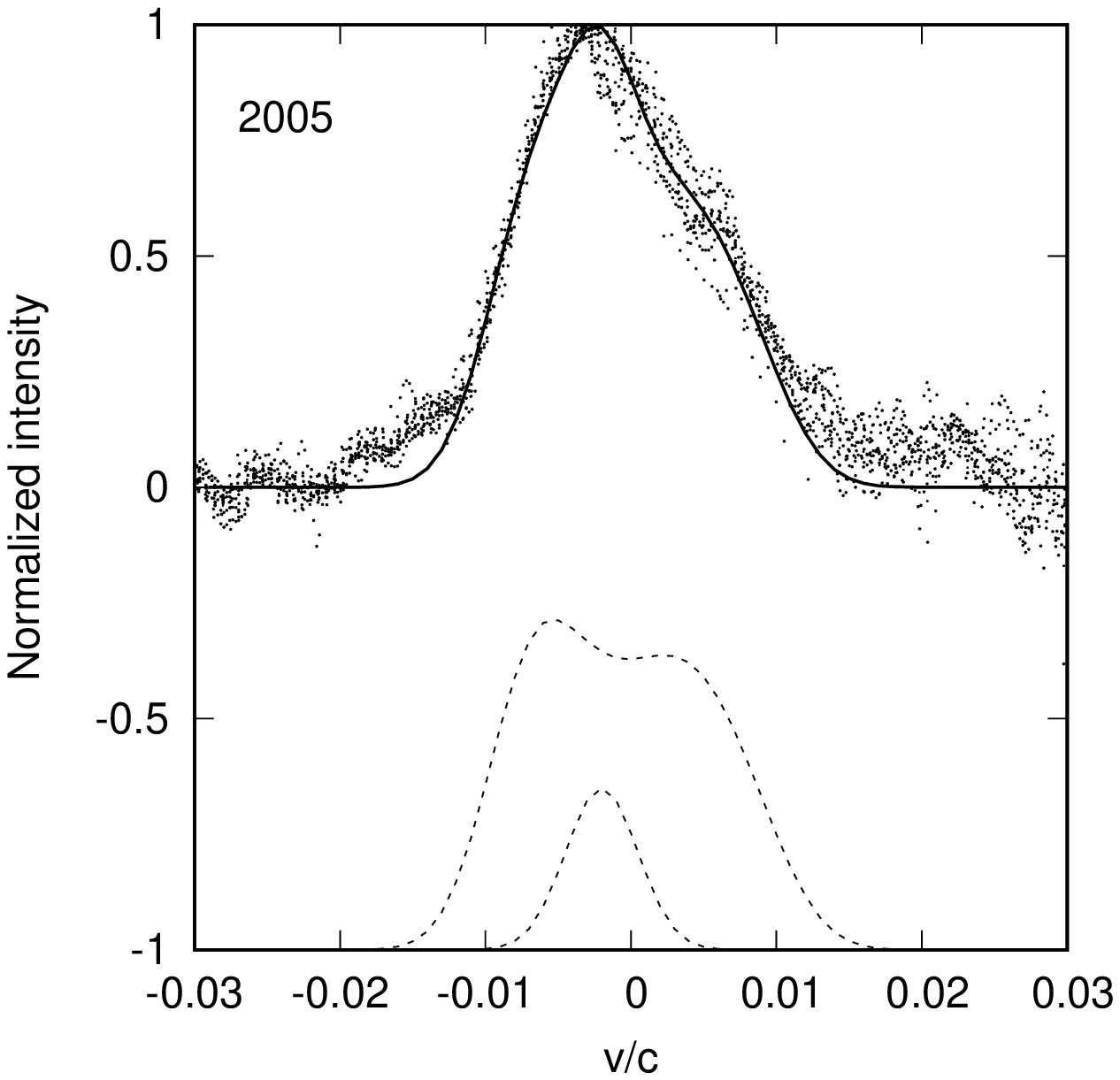}
\caption{The same as in Fig. \ref{fig-mod1} but for profiles observed in 2004, where minimum is given in second  and maximum in third panel. Forth panel is the model for H$\beta$ profile observed in 2005.}
\label{fig-mod4}
\end{figure*}

\begin{figure*}[]
\centering
\includegraphics[width=6cm]{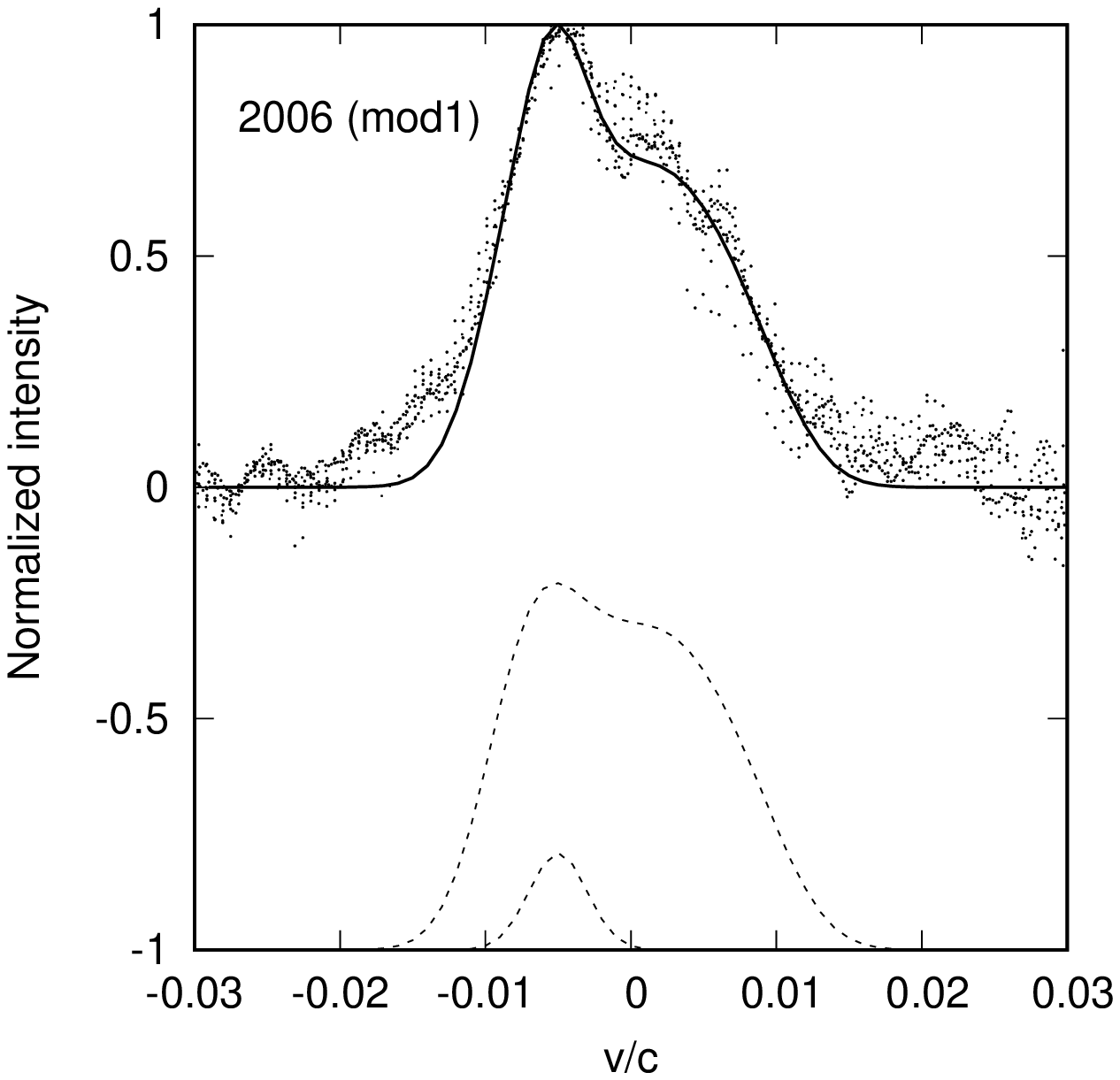}
\includegraphics[width=6cm]{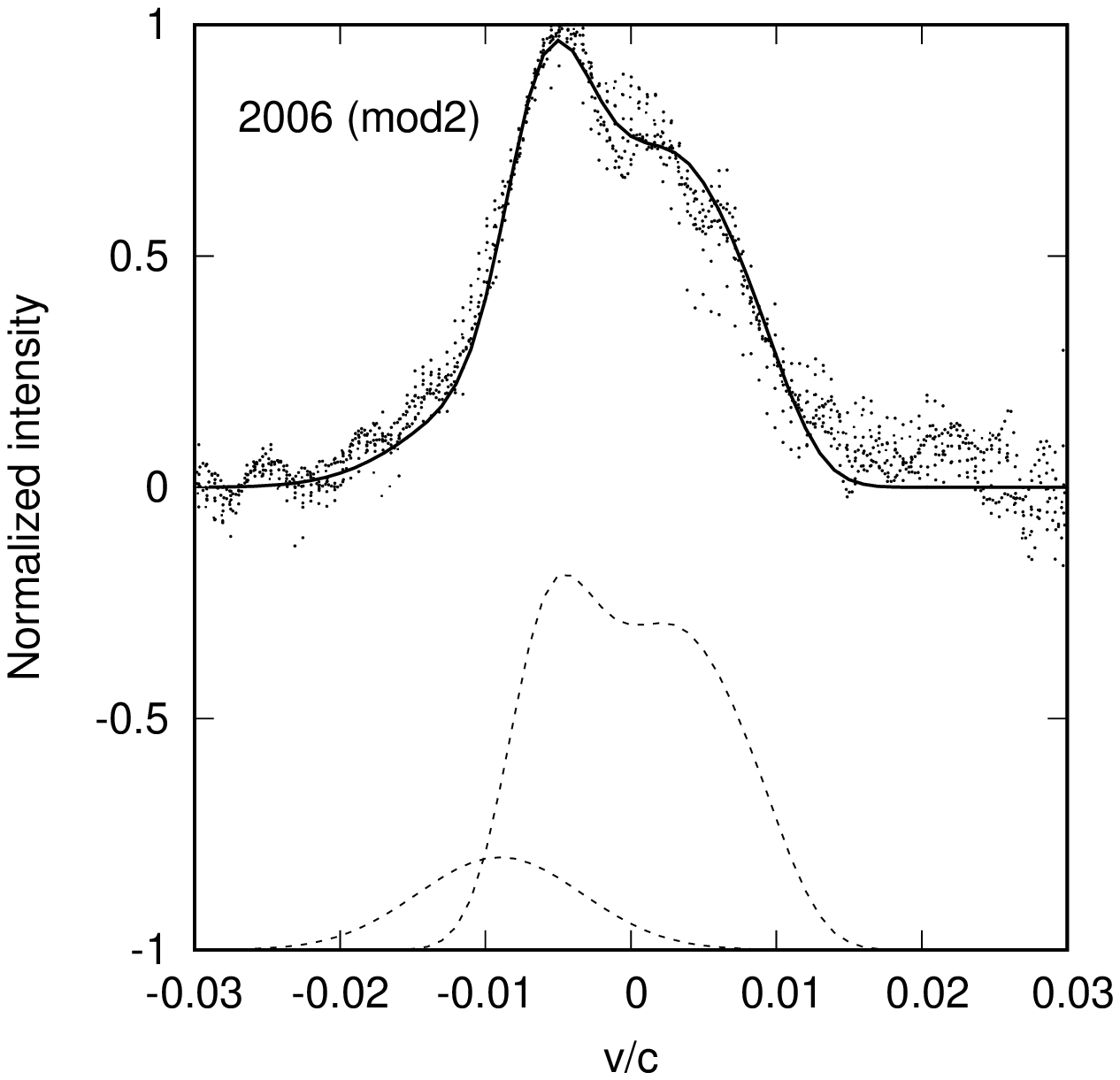}
\includegraphics[width=6cm]{fig_model_2007.eps}
\caption{The same as in Fig. \ref{fig-mod1} but for an averaged profile observed in 2006 and 2007, where line maximum intensity was observed.}
\label{fig-mod5}
\end{figure*}

\begin{figure*}[]
\centering
\includegraphics[width=8cm]{fig_model_2017.eps}
\includegraphics[width=8cm]{fig_model_2019.eps}
\includegraphics[width=8cm]{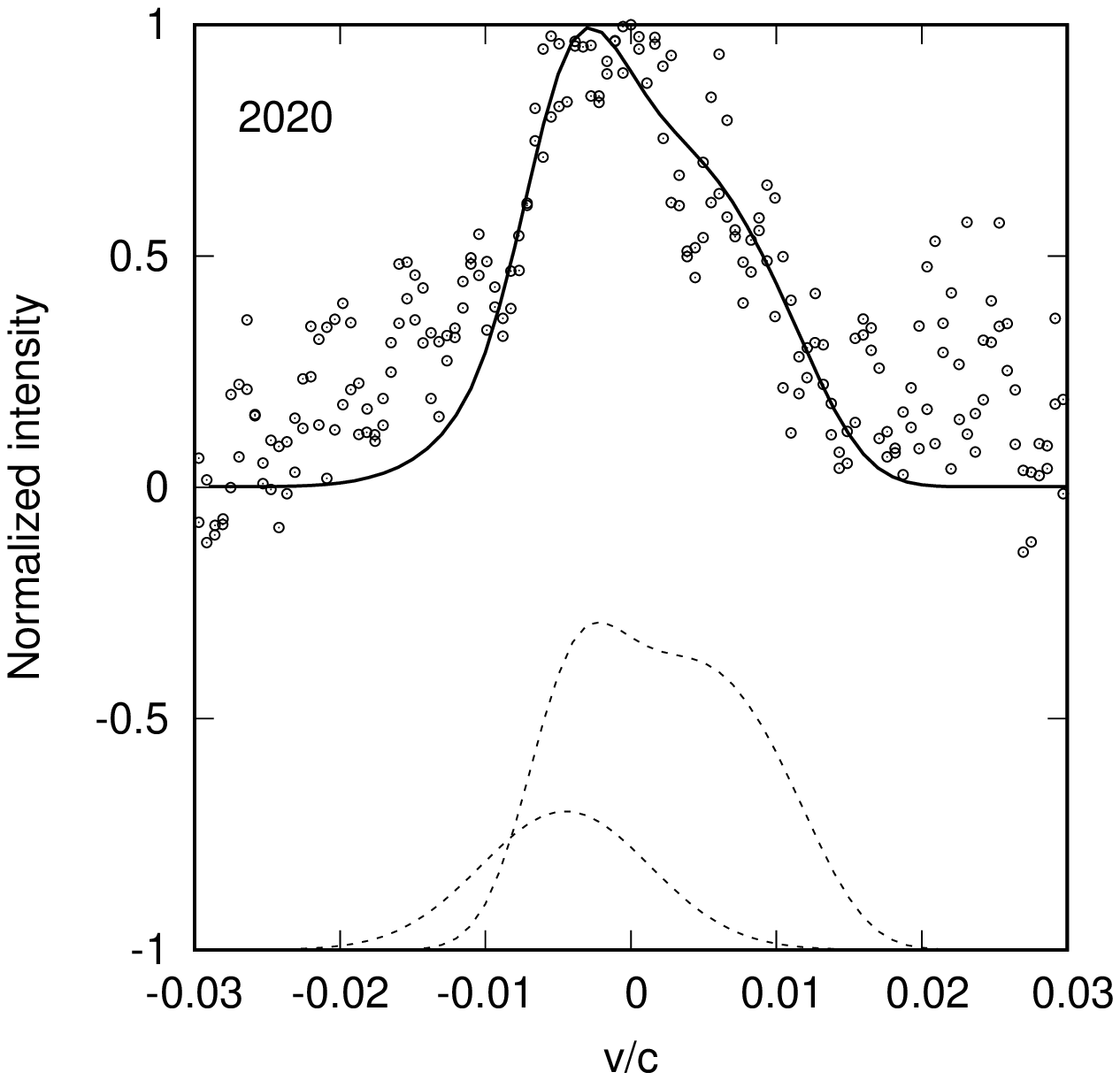}
\includegraphics[width=8cm]{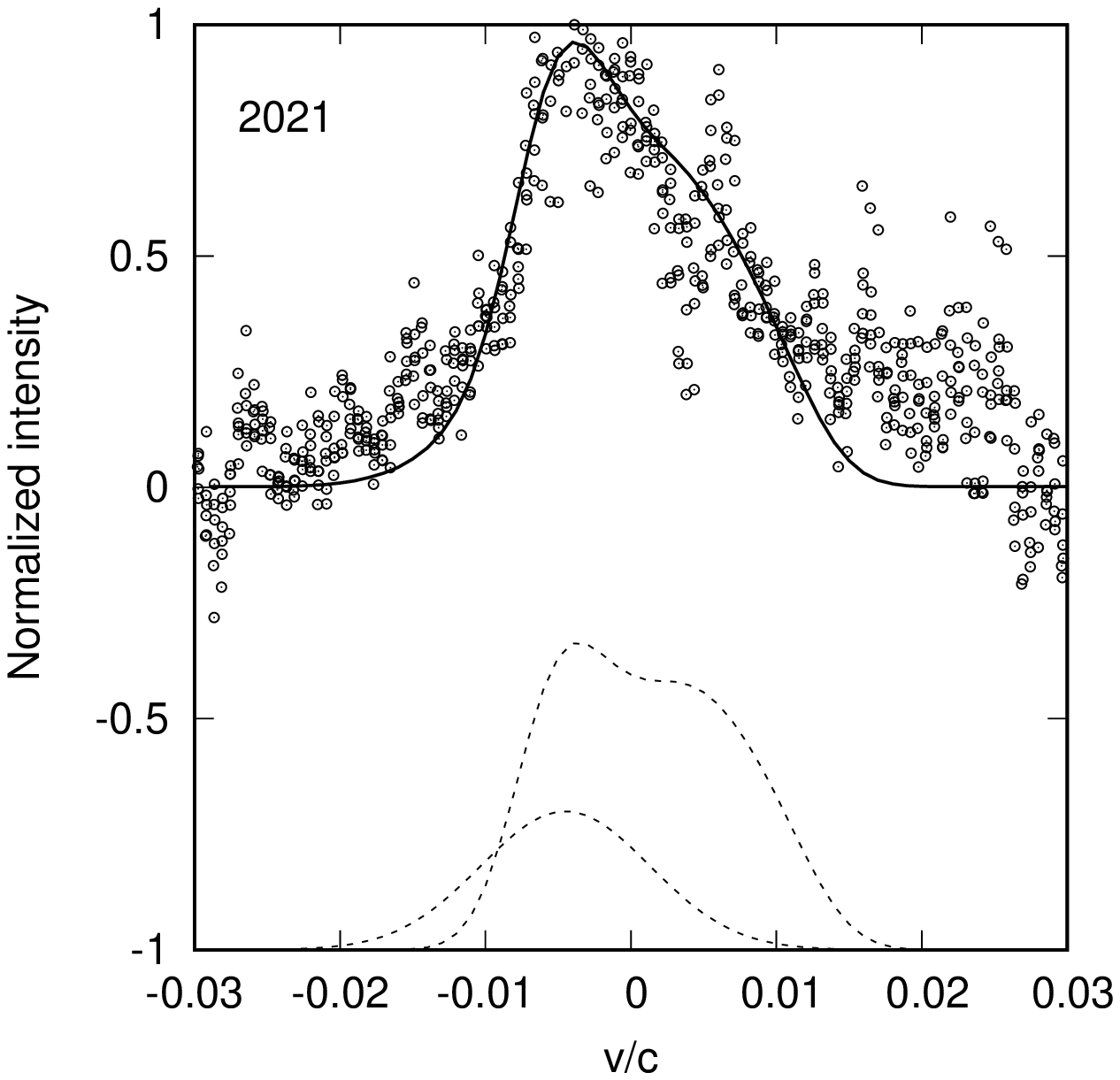}
\caption{The same as in Fig. \ref{fig-mod1} but for an averaged H$\alpha$ profiles observed in 2017,  2019 and for H$\beta$ observed in  2020 and 2021 where line minimum in broad lines was observed.}
\label{fig-mod6}
\end{figure*}

\end{document}